\documentclass[preprint,12pt,tightenlines,eqsecnum,floats,aps,amsmath,amssymb%
,nofootinbib,prd,showpacs]{revtex4}

\usepackage{amsmath,amssymb,amsfonts}
\usepackage{graphicx}
\usepackage{enumerate}

\usepackage{LQC-symbols}

\begin{document}

\preprint{\vbox{\baselineskip=12pt \rightline{IGC-08/3-3}
}}

\title{Anti-deSitter universe dynamics in LQC}

\author{Eloisa Bentivegna${}^{2,3}$}
\email{bentiveg@gravity.psu.edu}
\author{Tomasz Pawlowski${}^{1,2}$}
\email{tomasz@iem.cfmac.csic.es}
\affiliation{
  ${}^{1}$Instituto de Estructura de la Materia, \\
  Consejo Superior de Investigaciones Cient\'ificas (CSIC), \\ 
  Serrano 121, 28006 Madrid, Spain\\
  ${}^{2}$Institute for Gravitational Physics and Geometry,\\ 
  Physics Department, Penn State, \\
  University Park, PA 16802, U.S.A.\\
  ${}^{3}$Center for Gravitational Wave Physics,\\ 
  Physics Department, Penn State, \\
  University Park, PA 16802, U.S.A.
}

\begin{abstract}
  A model for a flat isotropic universe with a negative cosmological
  constant $\Lambda$ and a massless scalar field as sole matter content is
  studied within the framework of Loop Quantum Cosmology. By application
  of the methods introduced for the model with $\Lambda=0$, the physical
  Hilbert space and the set of Dirac observables are constructed. As in
  that case, the scalar field plays here the role of an emergent
  time. The properties of the system are found to be similar to those of the
  $k=1$ FRW model: for small energy densities, the quantum dynamics reproduces
  the classical one, whereas, due to modifications at near-Planckian
  densities, the big bang and big crunch singularities are replaced by a
  quantum bounce connecting deterministically the large semiclassical
  epochs. Thus in Loop Quantum Cosmology the evolution is
  qualitatively cyclic.
\end{abstract}


\maketitle

\section{Introduction}
\label{sec:intro}

Loop Quantum Cosmology \cite{lqc} -- an application of methods of Loop
Quantum Gravity \cite{lqg} to symmetry reduced models -- constitutes a
promising way of studying quantum-gravitational effects in
cosmological models. In particular one of the simplest models, a flat 
Friedmann-Robertson-Walker (FRW) universe was analyzed within its
framework \cite{aps,aps-old,aps-imp}. In that case, the structure of the
Hamiltonian constraint allowed to treat the constrained system as a
free one, evolving with respect to the scalar field which thus plays the
role of an emergent time. This, in turn, allowed the construction of
a physical Hilbert space and a set of Dirac observables, which were used 
next to extract the physics by means of numerical methods. The results were
quite surprising: the analysis has shown that, when the matter energy
density approaches the Planck scale, the quantum-geometric effects
cause gravity to become repulsive. In consequence, a large
semiclassical expanding universe is preceded by a (also large and
semiclassical) contracting one, deterministically connected to the
former by a quantum bridge. The transition point of the evolution 
(called quantum bounce) is characterized by an energy density which, at
this point, equals the critical value $\rho_c \approx 0.82\rhoPl$. 
Furthermore, even when quantum corrections actually dominate the
dynamics, the state representing the universe remains semiclassical --
its evolution is to great precision described by the so called
classical effective dynamics \cite{pv,aps-imp}.   

The results obtained for the flat FRW model were next generalized to
the spherical one \cite{apsv} (the $k=1$ FRW model). The properties of
the Hilbert space and an evolution operator were investigated analytically
\cite{klp,klp-nL} and the robustness of their features was confirmed
through the analysis of its approximation (known as {\it sLQC})
\cite{acs,cs-rec}. Further generalizations to anisotropic (and
further inhomogeneous) models by different research groups are in
various stages of progress \cite{chv,gp,gm-letter}. 

Thus far, however, the only models described rigorously were universes
with a vanishing cosmological constant $\Lambda$ and a massless
scalar field. In this article, we extend the analysis of \cite{aps-imp}
to include the universes with negative $\Lambda$.   
Although the observations favor a positive $\Lambda$, this model 
constitutes a convenient way of testing which features of the previously 
investigated model we can hope to generalize to more realistic
systems. Also, since it is a classically recollapsing system, we can use
it to investigate semiclassicality issues (dispersion after many 'cycles'
of evolution). The specific questions we intend to address here are the
following: 
\begin{itemize}
  \item Do the qualitative features of the $\Lambda=0$ model survive
    also in this case? In particular, are the big bang/crunch
    singularities replaced by quantum bounces as in the previously
    investigated cases? All the models analyzed so far not only
    experienced the bounce, but for Gaussian states the observed 
    dispersion of the wave packet after the bounce was severely restricted
    by the values of the spreads before it. In the flat case this result
    was next generalized analytically to a space of states admitting
    semiclassical epoch\footnote{The states for which either
      at early or late times the relative dispersions of chosen Dirac
      observables are $\ll 1$.} \cite{cs-rec} within the context of
    sLQC. Therefore, 
    it is important to ask whether such behavior will occur also in
    the considered model, or it was just a result of the extreme
    simplicity of the previous ones. 
  \item If the answer to the previous question is in the affirmative, then
    is the critical energy density $\rho_c$ still a fundamental bound? 
    In both the $k=0$ and $k=1$ models for physically 
    sensible\footnote{This indicates the states of the scalar field
      with momentum sufficiently high for the closed universe to grow
      to macroscopic ($>1$ megaparsec) scales before recollapsing.}
    states, the matter energy density at the bounce point agreed to
    great precision with $\rho_c$. Furthermore, later investigations
    within the sLQC model have shown that $\rho_c$ is indeed a
    fundamental energy bound. But again, we do not know a priori
    whether this feature is characteristic just of the models
    investigated so far and how (if at all) it generalizes.  
  \item Does this model possess any new feature not observed in
    $\Lambda=0$ or $k=1$ case ? 
\end{itemize}
A preliminary investigation of the $\Lambda<0$ model has been conducted
already in \cite{aps-imp}. However, the physical Hilbert space was not
constructed; the goal there was only to verify the persistence of the
bounce. Recently, a heuristically constructed effective classical
Hamiltonian was used \cite{eff} to obtain the effective trajectories
of both the $\Lambda<0$ and $\Lambda>0$ systems and analyze the effect
of the quantum-geometric corrections on the universe's dynamics. 
However, since the effective Hamiltonian was not derived systematically, 
the results have to be confirmed against genuine quantum evolution. 

In addition to the problems described above, we also address the concerns
about the choice of the symmetric sector of the physical Hilbert space
that is sometimes raised. Because of the absence of fermions, the triad
orientation reflection is a large gauge symmetry. This allowed one to
restrict the physical Hilbert space to the states symmetric under parity
reflection. However, since the choice of antisymmetric states 
is equally justified, it is natural to ask whether the results of LQC
are robust and will continue to hold if the antisymmetric sector 
is chosen. We address this issue by analyzing, in addition to the standard
symmetric states, also the space of antisymmetric ones and establish
robustness.  

The paper is organized as follows: 
we start with a brief summary of the basic framework (introduced
already in earlier papers) in section \ref{sec:frame}. Its content is
divided into three parts: the classical theory, the kinematics of LQC
and the derivation of the quantum Hamiltonian constraint.
In section \ref{sec:WDW} we consider a geometrodynamical equivalent of
the model -- the Wheeler-DeWitt (WDW) one. The reason for that is
two-fold: first, it will allow us to compare the results of LQC
against a standard quantum model and identify the nonperturbative
quantum-geometric effects. Second, it will serve as an introduction to
the methodology of extracting physics, used next on the LQC model. The
analytical solvability of the WDW model will allow us to 
show these methods without having to deal with the complications of
numerical analysis.
Analysis of the physical sector is carried out in section 
\ref{sec:LQC}. There, we extensively use the results of the numerical
study described in turn in section \ref{sec:num}. That section
contains also a description of the construction and analysis of the
states semiclassical at late times. The final results and their
discussion are placed in section \ref{sec:concl}.   

Apart from the main body, the article contains two appendices: in section
\ref{sec:antisymm}, we analyze the space of antisymmetric states,
whereas \ref{sec:heu} contains a description of the heuristic methods used
to extract some of the results.

\section{The LQC quantization scheme}
\label{sec:frame}

In this section, we introduce the quantization framework used in later
sections of the paper. Since we directly apply the framework described
in detail in \cite{aps-imp,apsv}, we will just present a brief sketch
of it. For a more detailed discussion, the reader is referred to the
above mentioned articles.  
  
The content of this section is divided into three parts. In the first, 
we present the classical theory used as a basis for quantization. 
The second part is dedicated to the description of the LQC kinematics. 
Finally, we recall the derivation of the LQC Hamiltonian constraint.

\subsection{Classical theory}
\label{sec:frameClass}

A flat ($k=0$) FRW model represents a spacetime admitting a foliation by
spatial isotropic $3$-surfaces $M$ of topology $\re^3$. Its metric
tensor can be written in the form 
\begin{equation}
  g = -\rd t^2 + a^2(t) \fidq \ ,
\end{equation}
where $t$ is a time parameter (the {\it cosmic time}), $\fidq$ is a unit
(fiducial) Cartesian metric on the surface $M$ and the function $a(t)$ is
called a scale factor. 

Due to the homogeneity and noncompactness of $M$, one cannot write an
action or Hamiltonian as an integral of the appropriate density over the
entire $M$. Instead, we can define them as integrals over a chosen fiducial
cubical cell $\fV$, constant in comoving coordinates\footnote{The considered 
  model is of the Bianchi type A: the equations of motion derived from
  the Hamiltonian specified in this way are identical to the Einstein
  field equations reduced to the isotropic case.}. Given such a cell,
one can define a triad $\fide$ (and cotriad $\fidw$ dual to it) as
directed along the edges of $\fV$ and orthonormal with respect to $\fidq$.

As gravitational phase space variables, we choose the connections
$A^i_a$ and the density-weighted triads $E^a_i$
\begin{subequations}\label{eq:AE-def}\begin{align}
  A^i_a\ &=\ c V_o^{-\frac{1}{3}}\,\fidw^i_a \ ,  &
  E^a_i\ &=\ p V_o^{-\frac{2}{3}}\sqrt{\fidq}\,\fide^a_i \ ,
  \tag{\ref{eq:AE-def}}
\end{align}\end{subequations}
where $V_o$ is a volume of $\fV$ with respect to $\fidq$. The real
parameters $c,p$ called respectively {\it connection} and {\it triad
  coefficients} 
coordinatize the ($2$-dimensional) phase space of the gravitational
degrees of freedom. Appropriate scaling by $V_o$ ensures the invariance of
the symplectic structure of this phase space (when written in terms of
$c,p$) under different choices of $\fidq$. The Poisson bracket
between $c$ and $p$ equals 
\begin{equation}
  \{c,p\}\ =\ \frac{8\pi\gamma G}{3} \ ,
\end{equation}
where $\gamma$ is the Barbero-Immirzi parameter.

The basic variables defined as in \eqref{eq:AE-def} automatically
satisfy the Gauss and diffeomorphism constraints. The contribution of
the geometry to the only nontrivial constraint -- the Hamiltonian
one -- is of the form 
\begin{equation}\label{eq:classHgrav}
  \Cgrav\ =\ -\frac{1}{\gamma^2} \int_{\fV} \rd^3 x \left( \varepsilon_{ijk}
             e^{-1} E^{ai} E^{bj} F^k_{ab} - \gamma^2\Lambda \right) 
        \ =\ -\frac{6}{\gamma^2} c^2\sqrt{p} 
             + {\Lambda}p^{\frac{3}{2}}
\end{equation}
where $e\ :=\ \sqrt{|\det E|}$ and the field strength
$F^k_{ab}\ :=\ 2\partial_a A^k_b + \varepsilon^k_{ij} A^i_a A^j_b$. 

The only matter content -- a homogeneous massless scalar field -- is
described by two global variables: the field value $\phi$ and its
conjugate momentum $p_{\phi}$, with Poisson bracket between them 
\begin{equation}
  \{\phi,p_{\phi}\}\ =\ 1 \ .
\end{equation} 
The pair $(\phi,p_{\phi})$ coordinatizes the phase space corresponding
to the matter degrees of freedom. The full phase space of the system
is thus $4$-dimensional. The complete Hamiltonian constraint is of the form
\begin{equation}\label{eq:classHcompl}
  C\ =:\ \Cgrav + \Cphi\ =\ 0 \ , \qquad 
  \text{where}\ \ \Cphi\ =\ 8\pi G p^{-\frac{3}{2}} p^2_{\phi} \ .
\end{equation}
The above constraint defines a $3$D hypersurface in the $4$D phase
space. Furthermore, since $C$ does not depend explicitly on $\phi$, the
momentum $p_{\phi}$ is a constant of motion. Therefore, the dynamical
trajectories can be represented as a (parametrized by $p_{\phi}$) family
of functions $p(\phi)$
\begin{equation}\label{eq:pre-class-traj}
  p(\phi)\ =\ \frac{(4\pi G)^{\frac{1}{3}}p_{\phi}^{\frac{2}{3}}}%
    {|\Lambda|^{\frac{1}{3}}\cosh(\sqrt{12\pi G}(\phi-\phi_o))}
\end{equation}
Their form implies that the considered system recollapses. Each
trajectory starts at the big bang singularity and ends in a big crunch.

\subsection{Kinematics of LQC}
\label{sec:frameKin}

To quantize the system, we follow the Dirac program. First we construct a
kinematical Hilbert space: in our case, it is the tensor product of
spaces corresponding to, respectively, gravitational and matter degrees
of freedom: $\Hilk = \Hilkg\otimes\Hilkf$. 

For the matter we apply the standard Schr{\"o}dinger quantization. 
As $\Hilkf$ we choose the standard Hilbert space of square integrable
functions $\Hilkf = L^2(\re,\rd\phi)$. The basic operators are
$\hat{\phi}$ and $\hat{p}_{\phi}$. To describe the state we choose the
(dual) basis $\dbra{\phi}$ of eigenstates of $\hat{\phi}$. The action
of $\hat{\phi}$, $\hat{p}_{\phi}$ on the state can be then expressed
as follows 
\begin{subequations}\label{eq:phiOp}\begin{align}
  \hat{\phi}\Psi(\phi)\ &=\ \phi\Psi(\phi) \ , &
  \hat{p}_{\phi}\Psi(\phi)\ &=\ -i\hbar\partial_{\phi}\Psi(\phi) \ , &
  \text{where}\ \Psi(\phi)\ :=\ \dip{\phi}{\Psi} \ .
\tag{\ref{eq:phiOp}}\end{align}\end{subequations}

The quantization of the gravitational degrees of freedom within LQC at the
kinematical level has been rigorously performed in \cite{abl}. The
procedure is the analog of the quantization scheme used in full LQG
(see for example \cite{lqc-MB}). 
Here the basic variables are triads and connections along straight
edges generated by $\fide^a_i$. The kinematical Hilbert space is
the space of square integrable functions on the Bohr compactification of
the real line $\Hilkg = L^2(\rBohr,\rd\mu_{\Bohr})$. We will represent
its elements using the basis consisting of the eigenfunctions of $p$ 
(promoted to an operator), labeled by $\mu\in\re$. Despite
the continuity of $\mu$, the elements of the chosen basis are
orthonormal with respect to Kronecker delta 
\begin{equation}\label{eq:kinIP}
  \sip{\mu_1}{\mu_2}\ =\ \delta_{\mu_1\mu_2} \ .
\end{equation}  
As basic quantum operators, we select $\hat{p}$ and $\whexp(i\fracs{\lambda
  c}{2})$ \footnote{Since the family $\whexp(i\lambda c/2)$ is not
  weakly continuous, the operator $\hat{c}$ does not exists.}. Their
action on the basis elements $\sket{\mu}$ is given by:   
\begin{subequations}\label{eq:ph-act}\begin{align}
  \hat{p}\sket{\mu}\ &=\ \frac{8\pi\gamma G\lPl^2}{6}\sket{\mu} \ , &
  \whexp(i\fracs{\lambda c}{2})\sket{\mu}\ &=\ \sket{\mu+\lambda}
  \ .
  \tag{\ref{eq:ph-act}}
\end{align}\end{subequations}
Since the holonomy along the edge of fiducial length $\lambda$
generated by $\fide^a_i$ can be expressed via $\exp(i\lambda c/2)$
\begin{equation}\label{eq:hexp}
  \hol{\lambda}{k}\ =\ \fracs{1}{2}[\exp(\fracs{i\lambda c}{2})
                        + \exp(-\fracs{i\lambda c}{2})]\id 
                  +\fracs{1}{i} [\exp(\fracs{i\lambda c}{2}) 
                        - \exp(-\fracs{i\lambda c}{2})] \tau_k 
\end{equation}
(where the $\tau_k$ are related to the Pauli matrices $\sigma_k$ via
$2i\tau_k=\sigma_k$), its quantum analog $\qhol{\lambda}{k}$ can
be expressed in terms of the operators $\whexp$ in the same way.

\subsection{LQC: the Hamiltonian constraint}
\label{sec:frameH}

In order to write the quantum operator corresponding to the Hamiltonian
constraint (\ref{eq:classHgrav},~\ref{eq:classHcompl}), we need to
reexpress it in terms of the basic objects selected in the previous
subsection.  

Let us start with $\Cgrav$ \eqref{eq:classHgrav}. The quantization of the
cosmological term is straightforward (and just amounts to promoting
$p$ to operator $\hat{p}$). The remaining part is an integral of the
product of two terms: $e^{-1} E^{ai} E^{bj}$ and $F^k_{ab}$. 

Following Thiemann \cite{ThTrick}, we can rewrite the first term 
in the following form
\begin{equation}\label{eq:eEE}
  \varepsilon_{ijk}e^{-1}E^{ai}E^{bj}\ =\ \sum_k
  \frac{\sgn(p)}{2\pi\gamma G\lambda V_o^{\frac{1}{3}}}
  \fidveps^{abc} \fidw^k_c \Tr \left( h^{(\lambda)}_k
  \{{h^{(\lambda)}_k}^{-1},V\}\tau_i \right)
\end{equation}
where $V=|p|^{\frac{3}{2}}$ is the (physical) volume of the cell $\fV$. 

The field strength term $F^k_{ab}$ can, on the other hand, be approximated
via holonomies along the square loop $\square_{ij}$ oriented on the
$i$-$j$ plane.  
\begin{subequations}\label{eq:Fdef}\begin{align}
  F^k_{ab}\ &=\ -2 
  \Tr\left( \frac{\hol{\lambda}{\square_{ij}}-1}{\lambda^2
    V_o^{\frac{2}{3}}} \right)
  \tau^k \fidw^i_a \fidw^j_b \ ,  &
  \hol{\lambda}{\square_{ij}}\ &=\ 
  \hol{\lambda}{i} \hol{\lambda}{j} \hol{\lambda}{i}^{-1}
  \hol{\lambda}{j}^{-1} \ .
  \tag{\ref{eq:Fdef}}
\end{align}\end{subequations}
The size of $\square_{ij}$ is fixed by the requirement that its
physical area equals the lowest nonzero eigenvalue of the LQG area
operator 
\begin{equation}\label{eq:mubar}
  \lambda\ =\ \bar{\mu}(\mu)\ \ \ \text{s.t.}\ \ \
  \Ar{}_{\square_{ij}}\ =\ \bar{\mu}^2|p|\ =\ \Delta\ 
  :=\ (2\sqrt{3}\pi\gamma)\lPl^2 \ .
\end{equation}
To express the action of the operator corresponding to $\hol{\bar{\mu}}{}$,
it is convenient to use, instead of the label $\mu$, a new label $v$ defined
as follows
\begin{equation}
  v\ :=\ K\sgn(\mu)|\mu|^{\frac{3}{2}} \ , 
  \qquad K\ :=\ \frac{2\sqrt{2}}{3\sqrt{3\sqrt{3}}} \ .
\end{equation}
In the new labeling an exponent operator
$\widehat{\exp}(\frac{i\bar{\mu}c}{2})$ --the component of
$\hol{\bar{\mu}}{}$ (via \eqref{eq:hexp})-- acts simply as a unit
translation 
\begin{equation}
  \widehat{\exp} ( \fracs{i}{2}{\bar{\mu}c} ) \sket{v}\ =\ \sket{v+1}
  \ .
\end{equation}

In the matter part of the Hamiltonian constraint, the only nontrivial
component is $|p|^{-3/2}$, but again this can be reexpressed in terms of
holonomies via Thiemann's method
\begin{equation}\label{eq:invVdef}
  |p|^{-\frac{3}{2}}\ =\ \sgn(p)\left[ \frac{1}{2\pi\lPl^2\gamma\bar{\mu}} 
                        \Tr\sum_k \tau^k h^{(\bar{\mu})}_k
                        \{ {h^{(\bar{\mu})}_k}^{-1} , V^{\frac{1}{3}}\}
                        \right]^3 \ .
\end{equation}

Finally, applying all the results (\ref{eq:hexp}-\ref{eq:invVdef}) to
\eqref{eq:classHcompl}, one can write the operator $\hat{C}$. We do so
choosing, in the process, a particular factor ordering (the so called
Kaminski ordering) \cite{aps-imp}, in which $\hCgrav$ is manifestly
symmetric and positive-definite. The action of the final result on
the state $\Psi\in\Hilk$ can be written in the following form
\begin{equation}\label{eq:Theta}
  \partial^2_{\phi}\Psi(v,\phi)\ =\ -\Theta\Psi(v,\phi)\ 
  =\ -\Theta_o\Psi(v,\phi) + [B(v)]^{-1} C_{\Lambda} \Psi(v,\phi)
  \ ,
\end{equation}
where $\Psi(v,\phi):=\sip{v,\phi}{\Psi}$ and the functions $B(v)$,
$C_{\Lambda}(v)$ equal
\begin{subequations}\label{eq:BCLcoeff}\begin{align}
  B(v)\ &:=\ \frac{27K}{8}|v|
             \left||v+1|^{\frac{1}{3}}-|v-1|^{\frac{1}{3}}\right|^3
             \ , &
  C_{\Lambda}(v)\ &:=\ \frac{16\pi^2\gamma^3\lPl^4}{27K\hbar} \Lambda
           |v| 
  \tag{\ref{eq:BCLcoeff}}
\end{align}\end{subequations}
and $\Theta_o$ is an operator corresponding to the $\Lambda=0$ case
derived in \cite{aps-imp} 
\begin{equation}\label{Theta0} 
  \Theta_o\Psi(v,\phi)\ 
  =\ -[B(v)]^{-1}\left( C^+(v)\Psi(v+4,\phi) + C^o(v)\Psi(v,\phi) 
         + C^-(v)\Psi(v-4,\phi) \right) \ ,
\end{equation}
with coefficients $C^{\pm}$, $C^o$, equal to
\begin{subequations}\label{eq:ThCoeffdef}\begin{align}
  C^+(v)\ &=\ \frac{3\pi KG}{8}|v+2|\, \big||v+3|-|v+1|\big| \ , & & \\
  C^-(v)\ &=\ C^+(v-4) \ , 
  &
  C^o(v)\ &=\ -C^+(v)-C^-(v) \ .
\end{align}\end{subequations}
For reasons we will explain in later sections of the paper, the operator
$\Theta$ is called an {\it evolution operator}. It is symmetric and
positive-definite (with respect to the measure $B(v)\rd\mu_{\Bohr}$) on
the domain $\Dom$ of finite linear combination of states $\sket{v}$.

\section{The Wheeler-DeWitt limit}
\label{sec:WDW}

The quantization scheme presented in the previous section is motivated by
LQG; however, it is not the only method applicable to the system.
By replacing $\Hilkg$ with $\ubHilkg:=L^2(\re,\rd\mu)$ and taking the
limit $\Delta\to 0$ in expressions 
(\ref{eq:eEE},~\ref{eq:Fdef},~\ref{eq:invVdef}),
one arrives to the system equivalent to the one originating from
geometrodynamics, known as the Wheeler-DeWitt system. In the literature, the
system obtained from LQC via this procedure is called a {\it WDW
  limit}. We will study it in this section in order to identify the
effects of the spacetime discreteness. We will keep this terminology
in the paper although (as it was shown in \cite{acs}) the WDW model
is not the limit of the LQC model in any precise sense. One
should think about it as the {\it WDW equivalent} of an LQC model.

\subsection{WDW constraint equation, emergent time}

The evolution operator $\Theta$ is a sum of two terms: a $\Lambda=0$
operator $\Theta_o$ and a $\Lambda$-dependent potential term \eqref{eq:Theta}. 
The WDW limit of $\Theta_o$ was derived in \cite{aps-imp} and is of
the form 
\begin{equation}
  \ub{\Theta}_o\,\ub{\Psi}(v,\phi)\ 
  = \ -12\pi G(v\partial_v)^2\ub{\Psi}(v,\phi)
\end{equation}
where $\ub{\Psi}\in\ubHilk:=\ubHilkg\otimes\Hilkf$. Calculating the
limit of the cosmological constant term requires just replacing $B$ in
the potential term by its point limit $\ub{B} := K/|v|$ for $\Delta\to 0$. 
In consequence, the WDW equivalent of equation \eqref{eq:Theta} has
the form  
\begin{equation}\label{eq:WDWmain}
  \partial_{\phi}^2\,\ub{\Psi}(v,\phi)\ 
  =\ -\ub\Theta\,\ub{\Psi}(v,\phi)\ 
  = 12\pi G(v\partial_v)^2\ub{\Psi}(v,\phi)
              +\frac{16\pi^2\gamma^3\lPl^4}{27K^2\hbar}\Lambda 
               v^2 \ub{\Psi}(v,\phi) \ ,
\end{equation}
where the operator $\ub{\Theta}$ is symmetric and positive-definite
with respect to the measure $\ub{B}\rd v$ in the standard domain of 
fast-decaying functions (Schwartz space). 

The above constraint divides the domain of $v$ into two independent
sectors, corresponding to different signs of $v$, i.e. to different
orientations of the triad $E^a_i$. Due to the absence of a parity
violating interaction in the considered system, we can restrict the 
studies to states that are symmetric/antisymmetric with respect to a reflection
in $v$. For further analysis, we choose the symmetric sector, that is 
$\ub{\Psi}(\phi,v)=\ub{\Psi}(\phi,-v)$; however, the presented
construction can be repeated directly also in the antisymmetric case, with
equivalent results. 

\subsection{General solutions, frequency decomposition}
\label{sec:WDWsols}

The constraint \eqref{eq:WDWmain} is similar in its form to the
Klein-Gordon equation. Furthermore, since there is no explicit
dependence on $\phi$ in either \eqref{eq:classHcompl} or
\eqref{eq:WDWmain}, $p_{\phi}$ is a constant of motion of both
the classical and the quantum system. Also, at the classical level $\phi$ is
monotonic in time: we can thus follow the prescription of
\cite{aps-imp} and reinterpret the constraint, treating it as an
evolution equation of a free system evolving with respect to
$\phi$. The scalar field becomes then an emergent time as in the case
$\Lambda=0$.  

To construct the physical Hilbert space we need to find the spectrum
of the self-adjoint extension of $\ub{\Theta}$. The eigenfunction
corresponding to an eigenvalue $\omega^2$ satisfying
\begin{equation}
  \omega^2\ub{\psi}(v)\ =\ -12\pi G(v\partial_v)^2\ub{\psi}(v)
              -\frac{16\pi^2\gamma^3\lPl^4}{27K^2\hbar}\Lambda 
               v^2 \ub{\psi}(v)
\end{equation}
can be written in terms of Bessel functions of the third kind
\begin{equation}\label{eq:WDWgeneig}
  \psi_{\omega}(v)\ =\ c_{(I)} \BesI_{ik}(\beta\sqrt{-\Lambda}|v|) 
            + c_{(K)} \BesK_{ik}(\beta\sqrt{-\Lambda}|v|)\ ,
\end{equation}
where $k := \omega/\sqrt{12\pi G}$, $\beta :=
2\sqrt{\pi\gamma^3}\hbar\lPl/(9K)$ and $c_{(I)},c_{(K)} \in \compl$.
When $\beta\sqrt{-\Lambda}|v|<k$, both $\BesI$ and $\BesK$ show
oscillatory behavior. In particular, as $|v|\to 0$, they approach
the eigenfunctions of the $\ub{\Theta}_o$ operator corresponding to the same
frequency $\omega$
\begin{equation}\label{eq:WDWelimit}
  \psi_{\omega}(v)\ =\ \tilde{c}^{+}\exp(ik\ln|v|)
    + \tilde{c}^{-}\exp(-ik\ln|v|) \ .
\end{equation}
The complex coefficients $\tilde{c}^{+}$, $\tilde{c}^{-}$ of the limit
can be determined uniquely as functions of $c_{(I)}$, $c_{(K)}$. 

For $\beta\sqrt{-\Lambda}|v|>k$, the functions $\BesI$ grow exponentially,
whereas the functions 
$\BesK$ exponentially decay. In consequence, only the eigenfunctions
with $c_{(I)}=0$ will contribute to the spectral decomposition of
$\ub{\Theta}$. This implies that the spectrum of $\ub{\Theta}$ equals
$\Sp(\ub{\Theta})=[0,\infty)$ and is continuous. Furthermore, due to
\eqref{eq:WDWelimit}, the eigenfunctions with $c_{(I)}=0$ are Dirac
delta normalizable. Therefore, we can choose the basis setting 
$\ub{e}_{\omega}:=\alpha(\omega)\BesK_{ik}(\beta\sqrt{-\Lambda}|v|)$,
where $\alpha$ is a real, positive, $\omega$-dependent normalization 
factor chosen to satisfy the relation
\begin{equation}
  \sip{\ub{e}_{\omega}}{\ub{e}_{\omega'}}\ =\ \delta(\omega,\omega') \ .
\end{equation}

At this point, we note that the structure of the spectral decomposition of
$\ub{\Theta}$ is similar to the one of the WDW limit for the $k=1$ FRW model
\cite{apsv}, so that we can follow the construction used there. Each
element $\psi(v)$ of $L^2(\re,\ub{B}(v)\rd v)$ can be decomposed in the
basis $\ub{e}_{\omega}$:
\begin{equation}
  \psi(v)\ =\ \int_0^{\infty} \rd \omega \tilde{\psi}(\omega)
             \ub{e}_{\omega}(v) \ .
\end{equation}
where $\tilde{\psi}\in L^2(\re,\rd\omega)$. Therefore, the solutions
to the evolution equation \eqref{eq:WDWmain} with initial data
in the Schwartz space can be represented in terms of the two functions 
$\Psi_{\pm}(\omega)\in L^2(\re,\rd\omega)$
\begin{equation}\label{eq:WDWgensol}
  \Psi(v,\phi)\ =\ \int\rd\omega\left[
    \tilde{\Psi}_+(\omega)\ub{e}_{\omega}(v)e^{i\omega\phi} +
    \tilde{\Psi}_-(\omega)\bar{\ub{e}}_{\omega}(v)e^{-i\omega\phi}
  \right] \ .
\end{equation}
The solutions with vanishing $\ub{\tilde{\Psi}}_{+}$ and
$\ub{\tilde{\Psi}}_{-}$ (denoted in the following as $\ub{\Psi}_-$,
$\ub{\Psi}_+$) are called the negative and positive frequency solutions
respectively. Their general form can be written in terms of the square
root of the $\ub{\Theta}$ operator; namely, for initial data $\psi_o(v)$
specified at $\phi=\phi_o$, we have: 
\begin{equation}\label{eq:WDWmainpm}
  \Psi_{\pm}(v,\phi)\ =\ e^{\pm i\sqrt{\ub{\Theta}}(\phi-\phi_o)}\psi_o(v) \ .
\end{equation}

\subsection{Physical Hilbert space, observables}
\label{sec:WDWphys}

To construct the physical Hilbert space $\ubHilp$, we again follow
\cite{aps-imp,apsv}. As $\ub{\Theta}$ is the sum of the $\ub{\Theta}_o$
operator (which is just $\partial_{\ln|v|}^2$) and the positive potential
term, it is essentially self-adjoint and positive-definite \cite{klp-nL}.  
Friedrich's extension of it is thus a unique self-adjoint one. One can
then apply group averaging techniques \cite{gave} (see the discussion in
\cite{aps-old}) to find $\ubHilp$ and the inner product. The result is the
following: the space $\ubHilp$ itself consists of normalizable
solutions to \eqref{eq:WDWmain}; however, as the spaces of positive 
and negative frequency solutions are superselected sectors, we can
take as $\ubHilp$ the restriction to just one of them. Following
previous works, we chose positive frequency part, thus defining
$\ubHilp$ as:  
\begin{equation}
  \ub{\Psi}(v,\phi)\ =\ \int\rd\omega\tilde{\ub{\Psi}}(\omega)
    \ub{e}_{\omega} e^{i\omega\phi} \ , \qquad \
    \tilde{\ub{\Psi}}\in L^2(\re^+,\rd\omega) \ .
\end{equation}
The physical inner product within this space can be written as:
\begin{equation}
  \sip{\ub{\Psi}}{\ub{\Phi}}\ 
  =\ \int_{\phi=\phi_o} \ub{B}(v)\rd v \bar{\ub{\Psi}}(v)\ub{\Phi}(v) \ .
\end{equation}

In order to be able to extract physical information out of our system,
we need to define a set of Dirac observables, i.e. self-adjoint operators
preserving $\ubHilp$. Here again we can directly use the scalar field
momentum $\hat{p}_{\phi}$ and $|\hat{v}|_{\phi}$ , the amplitude of $v$ at 
a given $\phi$, defined already for $\Lambda=0$ and $k=1$. Their 
action on the elements $\Psi$ of $\ubHilp$ is the following
\begin{subequations}\label{eq:WDWobs}\begin{align}
  \hat{p}_{\phi}\Psi\ &=\ -i\hbar\partial_{\phi}\Psi \ , & 
  |\hat{v}|_{\phi'}\Psi\ 
  &=\ e^{i\sqrt{\ub{\Theta}}(\phi-\phi')}|v|\Psi(v,\phi') \ ,
  \tag{\ref{eq:WDWobs}}
\end{align}\end{subequations}
and their expectation values equal respectively:
\begin{subequations}\label{eq:WDWexpect}\begin{align}
  \sbra{\ub{\Psi}}\hat{p}_{\phi}\sket{\ub{\Psi}}\ &= \
    -i\hbar\int_{\phi=\const}\ub{B}(v)\rd v \bar{\ub{\Psi}}(v,\phi)
     (\partial_{\phi}\ub{\Psi})(v,\phi) \ , \\
  \sbra{\ub{\Psi}}|\hat{v}|_{\phi}\sket{\ub{\Psi}}\ &= \
     \int\ub{B}(v)\rd v |v||\ub{\Psi}(v,\phi)|^2 \ .
\end{align}\end{subequations}

\subsection{Semiclassical states}
\label{sec:WDWstates}

Once we have the physical Hilbert space, the inner product and the 
observables, we
can investigate the evolution of a universe represented by a given
state. A particularly interesting question one can ask is
whether, in the considered system, the singularity is resolved. To address
this question, we construct a Gaussian state which, at a given time
$\phi_o$, is sharply peaked at a large scalar field momentum
$p_{\phi}^{\star}=\hbar\omega^{\star}$ (with spread $\sigma/\sqrt{2}$)
and volume $v^{\star}$ and is expanding: 
\begin{equation}\label{eq:WDWpsi}
  \Psi(v,\phi)\ =\ \int_0^{\infty}\rd\omega \,
    e^{-\frac{(\omega-\omega^{\star})^2}{2\sigma^2}} \,
    \ub{e}_{\omega}(v) \,
    e^{i\omega(\phi-\phi^{\star})} \ , 
\end{equation}
where 
\begin{equation}\label{eq:phistar}
  \phi^{\star}= \frac{1}{\sqrt{12\pi G}}\arcosh\left( 
    \frac{3K\sqrt{12\pi G}}%
    {(4\pi\gamma\lPl^2)^{3/2}} \frac{p_{\phi}^{\star}}%
    {\sqrt{|\Lambda|}v^{\star}} \right) + \phi_o \ .
\end{equation}

\begin{figure}[tbh!]\begin{center}
  $a)$\hspace{8cm}$b)$
  \includegraphics[width=3.2in,angle=0]{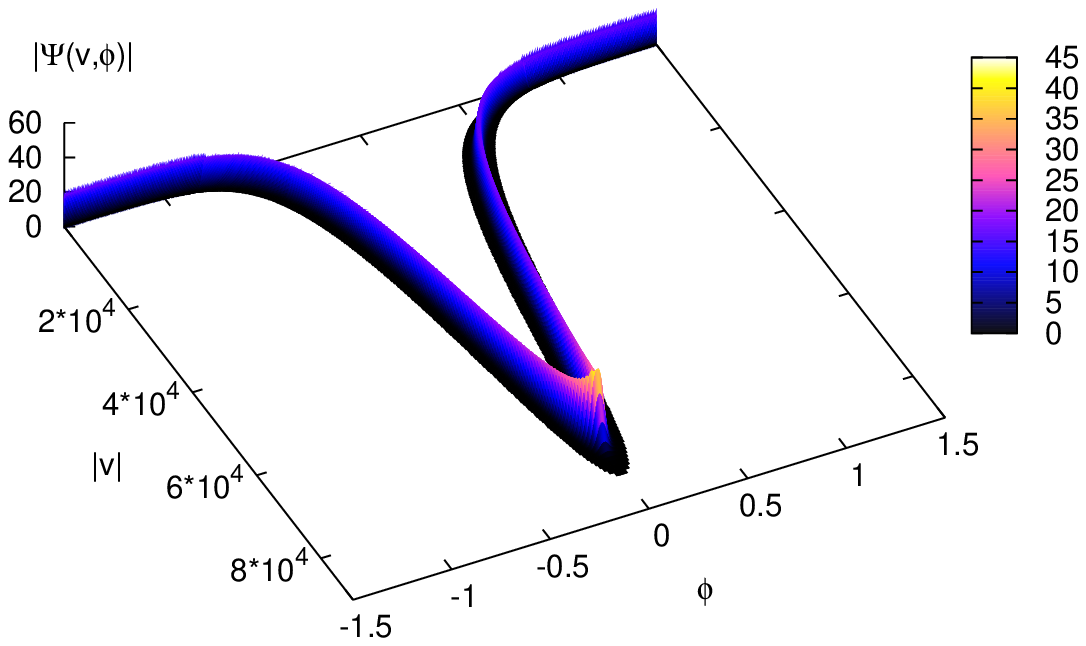}
  \includegraphics[width=3.2in,angle=0]{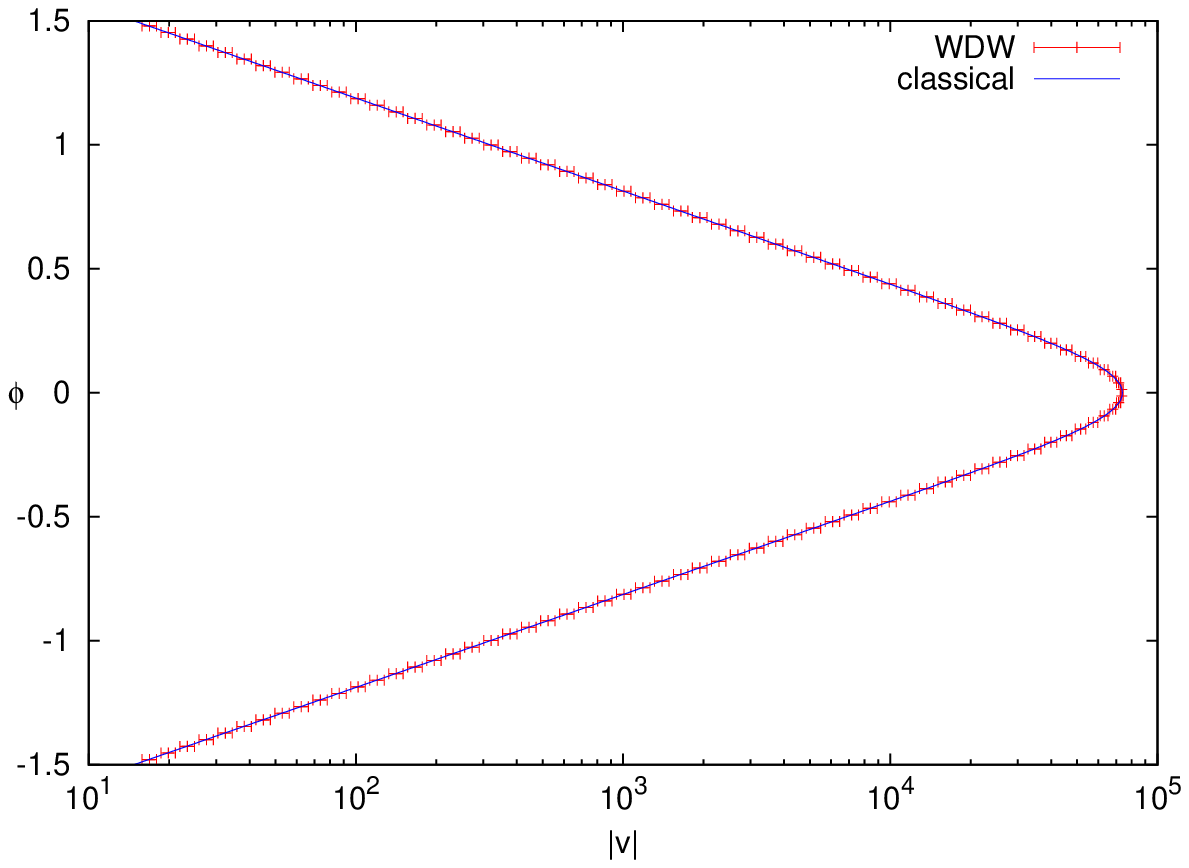}
  \caption{An example of a Wheeler-DeWitt Gaussian wave packet
    \eqref{eq:WDWpsi} generated for the parameter values
    $\Lambda=-0.01$, $p_{\phi}^{\star}=5\cdot 10^3$, $\Delta
    p_{\phi}/p_{\phi}^{\star}=0.02$ and $\phi^{\star}=0$. Fig.$a)$ shows the
    absolute value of the wave function. For the presentation clarity, only
    the points of $|\Psi(v,\phi)|>10^{-6}$ were plotted. Fig.$b)$
    presents the expectation values and dispersions of
    $|\hat{v}|_{\phi}$ (red bars) compared against the classical
    trajectory $v(\phi)$ (blue line). As we can see, the quantum
    trajectory agrees with the classical one (the difference being much
    smaller that the spread). Due to the large changes in magnitude of $v$
    during the evolution, the trajetory was plotted in logarythmic scale.} 
  \label{fig:WDW}
\end{center}\end{figure}

Because of the complicated form of $\ub{e}_{\omega}$, the wave function
\eqref{eq:WDWpsi} and expectation values \eqref{eq:WDWexpect} were
calculated numerically (see section \ref{sec:num} for the details). An
example of the results is shown on Figs \ref{fig:WDW}. The state remains
semiclassical (sharply peaked) and simply follows the classical trajectory
\eqref{eq:pre-class-traj}
\begin{equation}\label{eq:WDWtraj}
  v(\phi)\ =\ \frac{3K\sqrt{12\pi G}}{(4\pi\gamma\lPl^2)^{\frac{3}{2}}}
  \,\frac{p_{\phi}^{\star}}{\sqrt{|\Lambda|}}\, 
  \left[\cosh\left(\sqrt{12\pi G}(\phi-\phi^{\star}+\phi_o)
  \right)\right]^{-1} \ 
\end{equation}
to the big bang and big crunch singularities. In consequence,
similarly to the $\Lambda=0$ case, the classical singularities are not
resolved.  

\section{Physical sector of LQC}
\label{sec:LQC}

The analysis in the previous section allowed to find dynamics
predicted by the WDW limit of the considered LQC model. Now we perform
an analogous study of the model of interest. Due to qualitative
similarities of the Hamiltonian constraint with its WDW limit, the
analysis can be performed analogously to the one done in section
\ref{sec:WDW} (with only slight modifications required by the fact that
$\Theta$ is now a difference operator). Following that work, we again
restrict the study to states symmetric under parity
reflection.\footnote{It is also correct to work with the antisymmetric
  sector of the theory. We discuss that case in Appendix \ref{sec:antisymm}.} 

First we note that, thanks to the fact that $\Theta$ is a difference
operator, we can naturally divide the gravitational kinematical
Hilbert space onto superselected sectors $\Hilkg =
\bigoplus_{\varepsilon\in[0,2]} \Hilkge$, where $\Hilkge$ are the  
restrictions of $\Hilkg$ to the functions supported on the sets
$\lat_{\varepsilon} := \{\pm\varepsilon+4n;\ n\in\integ\}$ preserved by the
action of the Hamiltonian constraint \eqref{eq:Theta} and parity reflection 
$\Pi: \psi(v)\mapsto \psi(-v)$. Following the literature, we call these
sets {\it lattices} and work with single sectors 
$\Hilke:=\Hilkge\otimes\Hilkf$. The kinematical inner product
corresponding to them is just a restriction of the product of $\Hilk$. 

For each of the sectors illustrated above, the operator $\Theta$ is
obviously well defined and symmetric (with respect to the measure
$B(v)\rd\mu_{\Bohr}$) on the domain $D_{\varepsilon}$ -- the space of finite
combinations of $\sket{v}$ with $v\in\lat_{\varepsilon}$. Its
mathematical properties were rigorously analyzed in \cite{klp-nL}. It
is essentially self-adjoint, its extension is positive-definite and
its spectrum is discrete.  
The first two properties allow us again to choose $\phi$ as an emergent
time and treat $\Theta$ as an evolution operator. 

The discreteness of $\Theta$'s spectrum implies that the eigenfunctions
relevant for its spectral decomposition are normalizable. Furthermore,
a numerical study (discussed in section \ref{sec:num}) shows that
the spectrum is nondegenerate. In consequence, for each allowed value
of the label $\varepsilon$, we can build the physical Hilbert space
$\Hilpe$ as a space of normalizable positive frequency solutions to
\eqref{eq:Theta}, analogously to the construction in sections
\ref{sec:WDWsols} and \ref{sec:WDWphys}:
\begin{equation}\label{eq:Psi}
  i\partial_{\phi}\Psi\ =\ \sqrt{\Theta}\Psi \ , \qquad
  \Psi(v,\phi)\ =\ \sum_{n\in\natu} \tilde{\Psi}_n\, e_n(v)\, 
    e^{i\omega_n\phi} \ ,
\end{equation}
where $\Tilde{\Psi}$ are square summable and $e_n(v)$ are symmetric
in $v$ and normalized eigenfunctions of $\Theta$, corresponding to
eigenvalues $\omega_n^2$ which form the basis of $\Hilpe$.
The physical inner product can be found through group averaging
analogously to the WDW case and can be written in the form
\begin{equation}\label{eq:IP}
  \sip{\Psi}{\Phi}\ =\ \sum_{n=0}^{\infty} 
  \bar{\tilde{\Psi}}_n\tilde{\Phi}_n\
  =\ \sum_{v\in\lat_{\epsilon}} B(v)\bar{\Psi}(v)\Phi(v) \ .
\end{equation}

To complete the quantization program we need to choose a set of
Dirac observables. In order to be able to compare the results with the WDW
limit, we choose the operators analogous to \eqref{eq:WDWobs}
\begin{subequations}\label{eq:obs}\begin{align}
  \hat{p}_{\phi}\Psi\ &=\ -i\hbar\partial_{\phi}\Psi \ , & 
  |\hat{v}|_{\phi'}\Psi\ 
  &=\ e^{i\sqrt{\Theta}(\phi-\phi')}|v|\Psi(v,\phi') \ .
  \tag{\ref{eq:obs}}
\end{align}\end{subequations}
Their expectation values are equal respectively to
\begin{subequations}\label{eq:expect}\begin{align}
  \sbra{\Psi}\hat{p}_{\phi}\sket{\Psi}\ 
  &=\ -i\hbar
  \sum_{v\in\lat_{\epsilon},\ \phi=\const} B(v) 
        \bar{\Psi}(v,\phi) (\partial_{\phi}\Psi)(v,\phi) \ , \\
  \sbra{\Psi}|\hat{v}|_{\phi}\sket{\Psi}\
  &=\ 
  \sum_{v\in\lat_{\epsilon}} B(v) |\Psi(v,\phi)|^2 \ .
\end{align}\end{subequations}

To calculate an explicit form of $\Psi$ (needed to find the
expectation values) one needs to find the spectrum of $\Theta$ and the
explicit form of its normalizable eigenfunctions. Because of the
complicated structure of $\Theta$, in order to do so one needs to resort to
numerical methods. We present them in the next section.

\section{Numerical study}\label{sec:num}

This section is divided onto two parts. In section \ref{sec:num-eig},
we present the methods and results of identifying the spectrum of the
$\Theta$ operator and finding normalizable eigenfunctions. The
techniques for computing the wave function and the expectation values
are presented in section \ref{sec:num-state}. In both parts, we
applied the (appropriately refined) methods used already for the $k=1$
model and introduced in \cite{apsv}. Unless specified otherwise, from
now on we will work with units in which $G=1$. 
 
\subsection{Spectrum of $\Theta$}\label{sec:num-eig}
  
In order to construct the Hilbert space $\Hilpe$, one needs to find the
eigenfunctions supported on the lattice $\lat_{\varepsilon}$, which
consists of two sublattices
$\lat_{\pm|\varepsilon|}:=\{\pm|\varepsilon|+4n;\ n\in\integ\}$  
invariant with respect to the action of the Hamiltonian constraint. Each of
such eigenfunctions (denoted here as $\psi$) is a solution to a
difference equation: 
\begin{equation}\label{eq:eigen}
  -\omega^2 B(v)\psi(v)\ =\ C^+(v)\psi(v+4) 
    + (C^o(v)+C_{\Lambda}(v))\psi(v) + C^-(v)\psi(v-4) \ ,
\end{equation}
where $\omega^2$ is the eigenvalue that each given eigenfunction
corresponds to and $C^o, C^{\pm}, C_{\Lambda}$ are given by
(\ref{eq:BCLcoeff},~\ref{eq:ThCoeffdef}). On each sublattice, this is a
second-order equation -- 
one needs to specify the initial data at two neighboring points ($v_{\init}$,
$v_{\init}+4$) to uniquely define a solution. The symmetry condition
$\psi(v)=\psi(-v)$, however, restricts the amount of initial data in 
the following way:
\begin{enumerate}[(i)]
  \item For $\varepsilon\in (0,2)$, the sublattices $\lat_{\pm|\varepsilon|}$ 
    are disjoint and the parity reflection $\Pi$ transforms one onto another.
    Therefore one needs to specify an initial data $\psi(v_{\init})$,
    $\psi(v_{\init}+4)$ for just one of them, say
    $\lat_{+|\varepsilon|}$, and complete it by the action of $\Pi$. We denote
    such lattices as {\it generic}.  
  \item When $\varepsilon=0,2$ the sublattice $\lat_{+|\epsilon|}$
    coincides with $\lat_{-|\epsilon|}$ and is invariant with respect
    to parity reflection. The condition $\psi(v)=\psi(-v)$, applied to
    \eqref{eq:eigen}, imposes on it an additional constraint of the
    form depending on the value of $\varepsilon$: 
    \begin{itemize}
      \item $\varepsilon=0$: $\psi(-4)=\psi(0)=\psi(4)$,
      \item $\varepsilon=2$: $\psi(-2)=\psi(2)$.  Here the equality 
        $C^-(2)=C^+(-2)=0$ implies additionally $\psi(\pm 6) = 
        -[(\omega^2B(2)+C_{\Lambda}(2)+C^o(2))/C^+(2)]\psi(\pm 2)$.
    \end{itemize} In consequence, the value of $\psi$ at just one point
    ($v=0$ or $v=2$) determines the entire eigenfunction. These cases are
    called {\it exceptional}.
\end{enumerate}

The degrees of freedom specified above are complex; however, since
the coefficients of \eqref{eq:eigen} are real, $\psi$ satisfies it
iff so do its components $\Re(\psi)$, $\Im(\psi)$. Therefore, we can
safely restrict our study to a real $\psi$.  

Upon this restriction, the space of solutions to \eqref{eq:eigen}
is $1$-dimensional for exceptional lattices and $2$-dimensional for
generic ones. Once the initial data are specified appropriately for
each case, the function $\psi$ can be found by solving \eqref{eq:eigen}
iteratively. 

\begin{figure}[tbh!]\begin{center}
  $a)$\hspace{8cm}$b)$
  \includegraphics[width=3.2in,angle=0]{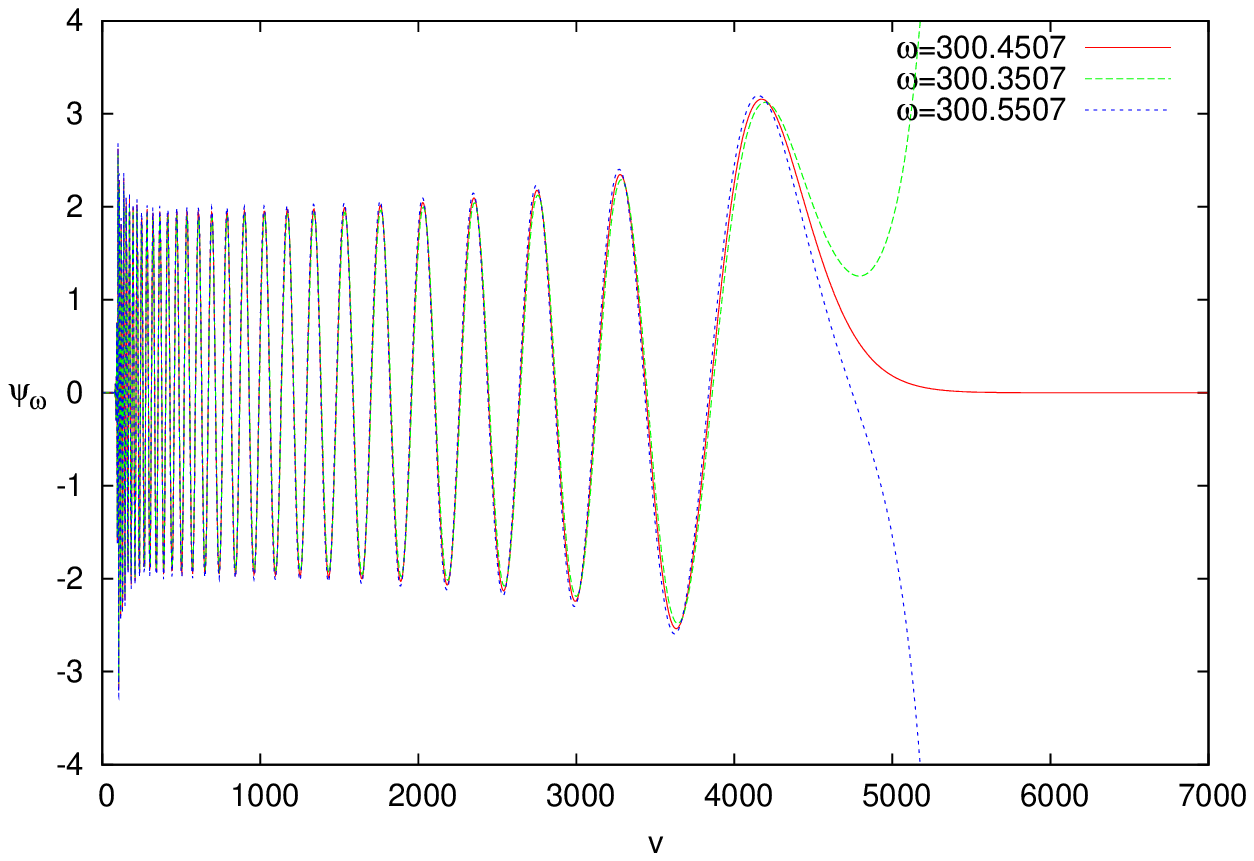}
  \includegraphics[width=3.2in,angle=0]{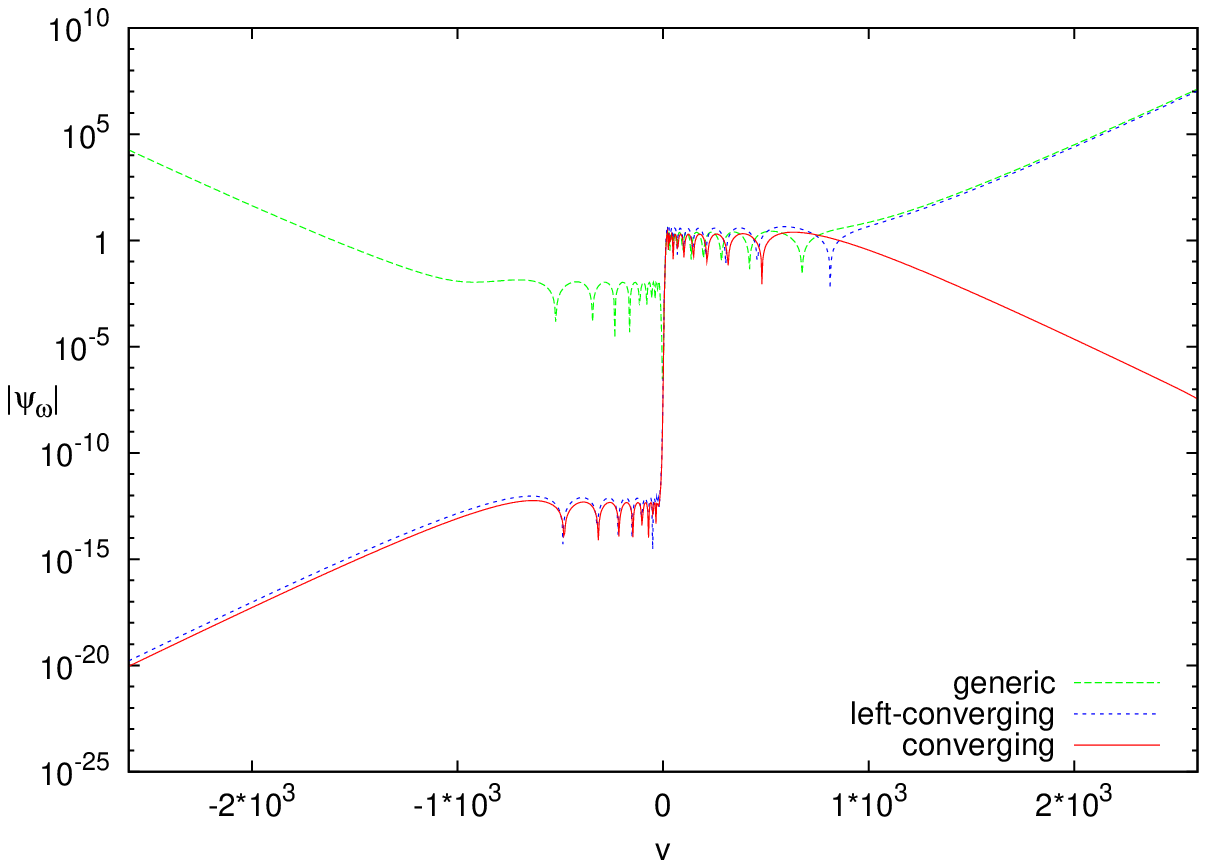}
  \caption{Examples of eigenfunctions of $\Theta$ supported on the
    lattices $\lat_{\varepsilon}$ for $\varepsilon=0$ $(a)$ and $\varepsilon=1$
    $(b)$.\\  
    $a)$ shows a normalizable eigenfunction of $\omega\approx300.45$
    (red) and two divergent ones of $\omega$ respectively smaller
    (green) and larger (blue) by $0.1$. For clarity, only the positive $v$
    part is shown.\\
    $b)$ presents the absolute value of a normalizable eigenfunction of
    $\omega\approx52.85$ (red) along with two divergent examples: generic
    (green) and left-converging (blue) generated for, respectively
    $\omega\approx53.35$ and $54.35$. To show the behavior in a wide range
    of values, a logarythmic scale was used for the $y$-axis.\\
    Both figures correspond to $\Lambda=-0.01$.}
  \label{fig:eig-exampl}
\end{center}\end{figure}

\begin{figure}[tbh!]\begin{center}
  $a)$\hspace{8cm}$b)$
  \includegraphics[width=3.2in,angle=0]{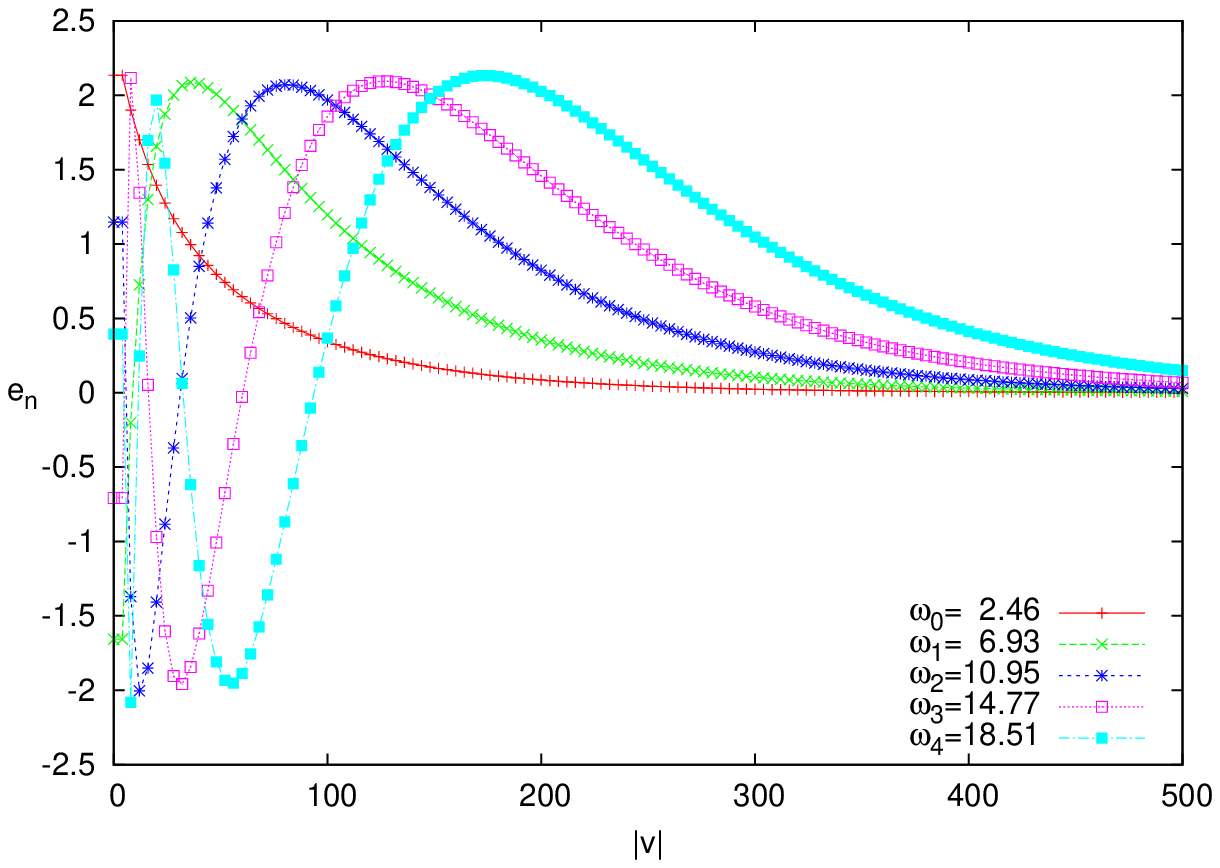}
  \includegraphics[width=3.2in,angle=0]{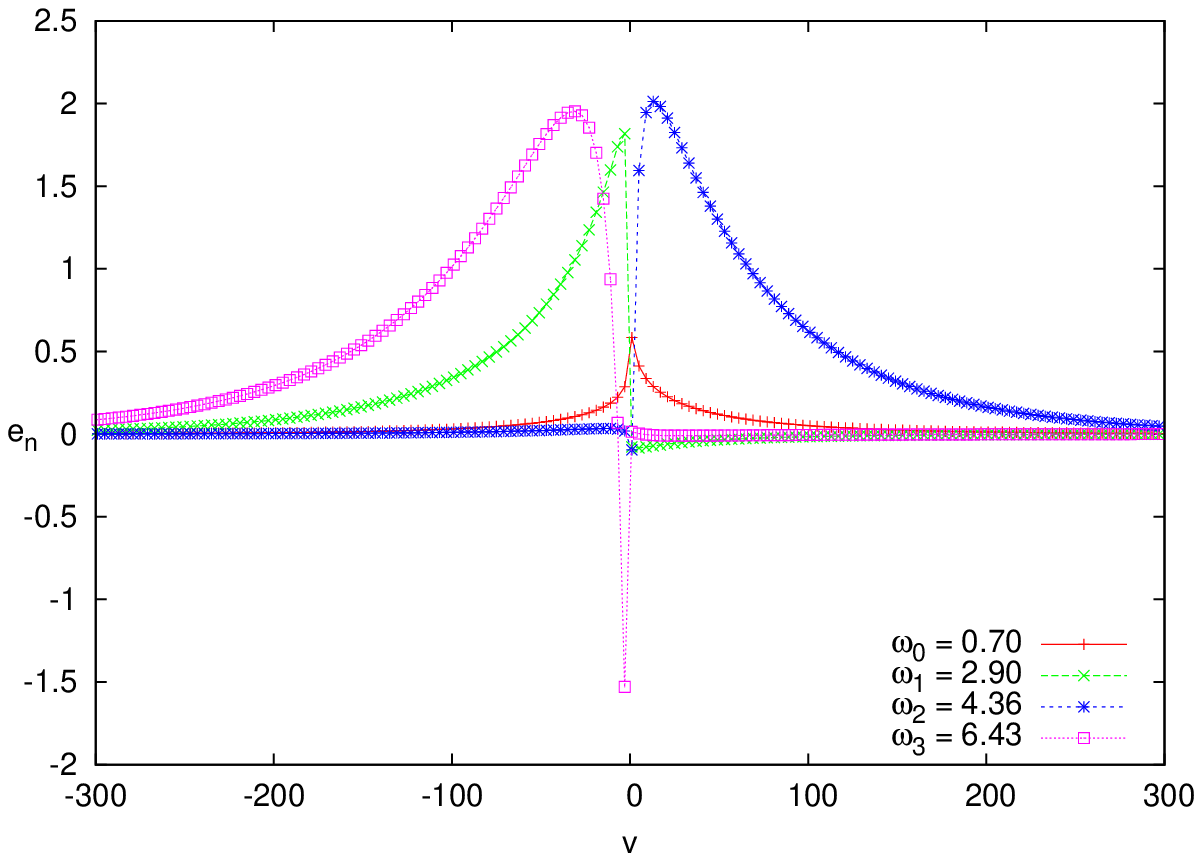}
  \caption{The eigenfunctions $e_0$ to $e_4$ of $\Theta$, corresponding to
    $\Lambda=-0.01$ and supported on $\lat_{\varepsilon}$ with
    $\varepsilon=0$ $(a)$ and $\varepsilon=1$ $(b)$. For clarity, only
    the $v>0$ part was shown in $a)$ and only the part supported on
    $\lat_{+|\varepsilon|}$ was shown in $b)$.}
  \label{fig:eig-low}
\end{center}\end{figure}

To determine the properties of $\psi$, we calculated the solutions in
a wide range of both $\Lambda$ ($[-10,-10^{-6}]$) and $\omega$
($[0,10^5\hbar]$). The qualitative features of the found solutions is 
visualized on fig.\ref{fig:eig-exampl}; in general, for each $\psi$ one
can distinguish $5$ zones of distinct behavior, and the boundaries of
these zones are specified by the functions $v_B(\omega)$ and $v_R(\omega)$,
approximately equal to, respectively, the position of the bounce for a
$\Lambda=0$ universe with $p_{\phi}^{\star} = \hbar\omega$ (determined in
\cite{aps-imp}) and the value of $v$ at the recollapse point of the classical
universe (given by \eqref{eq:WDWtraj} at $\phi=\phi^{\star}-\phi_o$).
\begin{enumerate}[(i)]
  \item For $|v|<v_B(\omega)$, the amplitude of $\psi$ grows/decays
    quasi-exponentially. \label{it:z1}
  \item For $v_B<|v|<v_R$, the behavior of $\psi$ is oscillatory
    (similar in nature to the behavior of \eqref{eq:WDWgeneig}). \label{it:z2}
  \item When $|v|>v_R$ the eigenfunction grows/decays exponentially
    with $|v|$ (where the exponential growth is a generic behavior).
    \label{it:z3}
\end{enumerate}
Note that for small $\omega$, the zones \eqref{it:z1} and
\eqref{it:z2} may be empty (see fig.\ref{fig:eig-low} for examples). 

Since we search for normalizable functions only, we have to select the
ones which decay exponentially in the zones of type \eqref{it:z3}. We
identify them numerically using different methods depending on whether 
the eigenfunctions are supported on generic or exceptional lattices.  

On exceptional lattices, each eigenfunction $\psi_{\omega}$ is (for
a given $\omega$) determined uniquely up to a global scaling. To find the
normalizable solutions, we scan the domain of $\omega$ using the
following observation: 

\begin{obs}
  For a chosen $\omega\in [\omega_{1},\omega_{2}]$,
  $\psi_{\omega}(\epsilon)=1$ and $v\gg v_R(\omega_2)$, the value
  $\psi_{\omega}(v)$ is a continuous function of $\omega$ (more specifically,
  a polynomial) and its sign changes
  quasi-periodically. Furthermore, if we define $\omega_{v,n}$ as the
  values of $\omega$ such that $\Psi_{\omega_{v,n}}(v) = 0$, the limits
  $\omega_n := \lim_{v\to\infty}\omega_{v,n}$ are well defined and
  correspond to the values of $\omega$ for which $\psi_{\omega}$
  decays in zone \eqref{it:z3}. 
\end{obs}

In practice, due to the precision bound posed by numerical round-off, 
it is enough to (instead of finding the limits) look for values
of $\omega_{v,n}$ at $v_T$ sufficiently far away from $v_R$. For the
actual search, we selected $v_T=\max(2000,1.3v_R)$. The search itself
was performed in two steps:
\begin{itemize}
  \item First the sign of $\Psi_{\omega}(v_T)$ was checked for
    values of $\omega$ uniformly separated by a distance around $0.1$.
  \item If a change of sign was detected between neighboring points, the
    value of $\omega_{n,v_T}$ was found via bisection. 
\end{itemize}

For generic lattices, the space of solutions is, up a to global rescaling,
$1$-dimensional, so besides $\omega$ we need to specify the value of $\psi$
at two points $v_I,v_I+4\in\lat_{+|\epsilon|}$. An additional
complication is the fact that now the behavior in zones of type \eqref{it:z3}
for $v>0$ and $v<0$ is independent. The function may grow for positive $v$
while decaying for negative ones and vice versa. Therefore, to find the
desired functions we divide the search procedure onto two steps:
\begin{itemize}
  \item First we identify the family $\psi_{\omega}$ of functions
    decaying in zone \eqref{it:z3} for $v<0$ (further denoted as {\it
      left-converging}). To do so, we parametrize the initial data at
    $v_I,v_I+4$ by a parameter $\alpha\in[0,\pi]$ 
    \begin{subequations}\label{eq:ida}\begin{align}
      \psi_{\alpha,\omega}(v_I)\ &=\ \cos(\alpha) \ , &
      \psi_{\alpha,\omega}(v_I+4)\ &=\ \sin(\alpha) \ ,
    \tag{\ref{eq:ida}}\end{align}\end{subequations}
    and scan the domain of $\alpha$ for the values at which the limit  
    $\lim_{v\to-\infty}\psi_{\alpha,\omega}(v) = 0$. Analogously to
    the exceptional lattice case, it is enough here to just choose some
    value $-v_R(\omega)\gg v_{T^-}\in\lat_{+|\varepsilon|}$ and look
    for the values of $\alpha$ at which $\psi_{\alpha,\omega}(v_{T^-})=0$.
    In practice, it suffices to choose $v_{T^-}\approx -v_T$, where
    $v_T$ is the value defined for exceptional lattices. The scan
    method is analogous to the scan of $\omega$ in the exceptional case: we
    divide the domain of $\alpha$ into $10$ uniform intervals and if
    a change of sign of $\psi_{\alpha,\omega}(v_{T^-})$ is detected
    within an interval, the precise value of $\alpha$ is found via
    bisection. 

    It was checked by inspection that, for each $\omega$, there is
    exactly one value of $\alpha$ satisfying the above requirement. In
    consequence, for each $\omega$ the eigenspace of left-converging
    functions is $1$-dimensional.
  \item Once the family $\psi_{\omega}$ of left-converging functions
    is selected, we choose some $v_T\approx
    v_{T^+}\in\lat_{+|\varepsilon|}$ and scan the domain of $\omega$ 
    for values at which $\psi_{\omega}(v_{T^+})=0$, via the method
    specified for exceptional lattices.
\end{itemize}

\begin{figure}[tbh!]\begin{center}
  $a)$\hspace{8cm}$b)$
  \includegraphics[width=3.2in,angle=0]{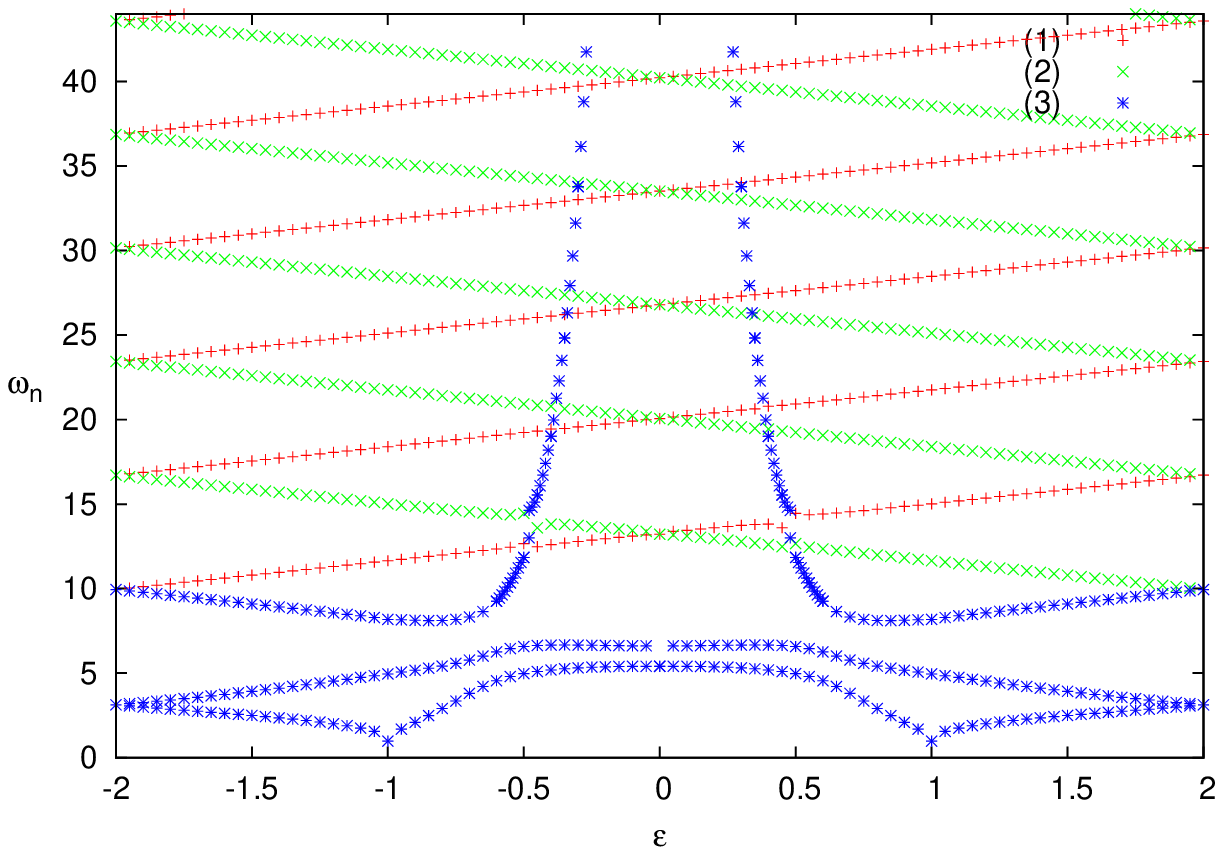}
  \includegraphics[width=3.2in,angle=0]{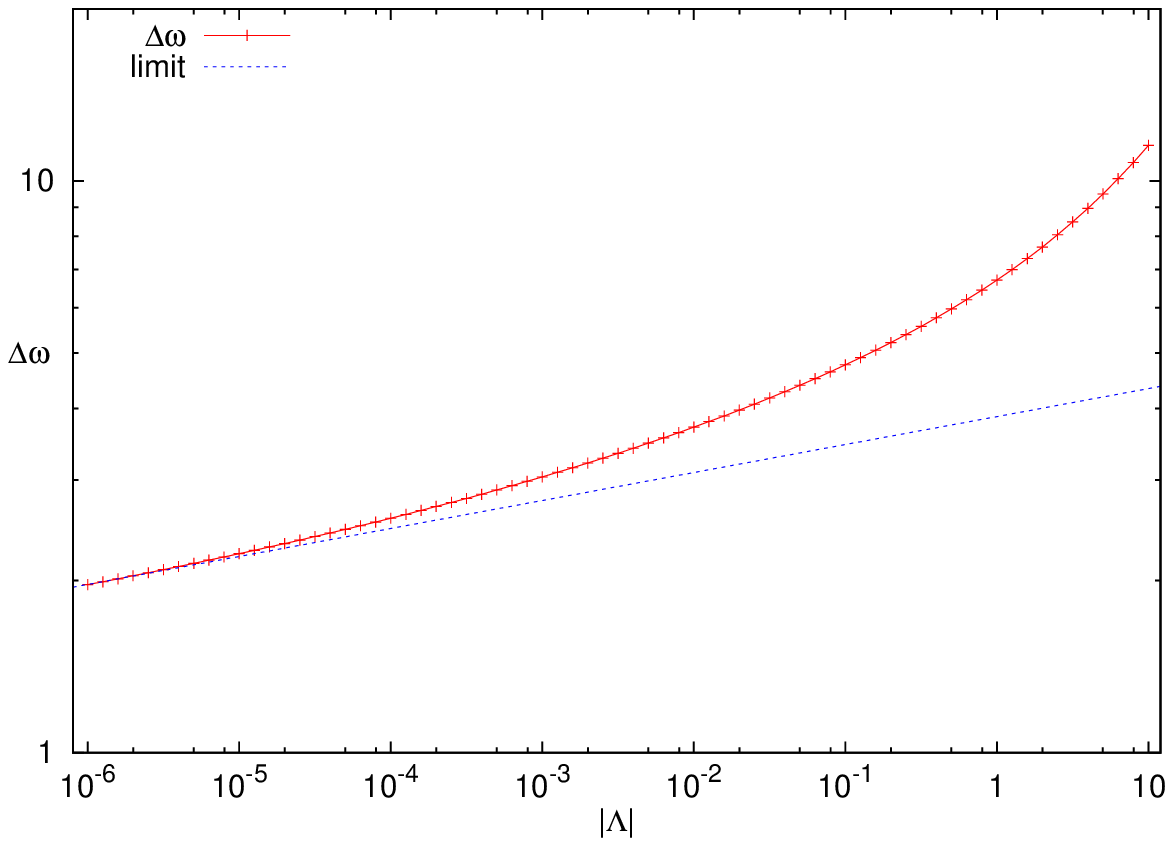}
  \caption{$a)$ The lowest ($\omega<44$) elements of $\Theta$'s spectrum are
    shown as functions of $\pm|\varepsilon|$. The eigenvalues are divided
    into three groups corresponding to the following eigenfuctions:
    $(1)$ left-suppressed (red crosses), for which $>50\%$ of 
    the norm is located on $v>0$, $(2)$ right-suppressed (green x-es)
    defined analogously and $(3)$ singularity-peaked (blue stars),
    where $>50\%$ of the norm is located at the three points closest
    to $v=0$. \\ 
    $b)$ The large $\omega$ limit of the eigenvalue separation
    $\Delta\omega$, shown as function of $\Lambda$ (red crosses). The
    blue line represents the small $\Lambda$ limit given by \eqref{eq:dwlow}.}
  \label{fig:dw}
\end{center}\end{figure}

The search was first performed for small $\omega$ ($<50$) to find the
qualitative behavior of normalizable eigenfunctions. An example of the
results is shown in Figs.~\ref{fig:eig-low} and \ref{fig:dw}a. All
found eigenfunctions belong to one of the following groups:
\begin{enumerate}[(1)]
  \item Suppressed on the $v<0$ side with suppression exponential in
    $\omega$. 
  \item Suppressed for $v>0$.
  \item Peaked about $v=0$.
\end{enumerate}
In consequence, it is most convenient, from the point of view of
the numerical precision of the solutions, to specify the initial data at 
$v_I\approx\pm v_R$. However, because of the quasi-exponential behavior of
the eigenfunctions in zone \eqref{it:z1},  we can calculate (with a sufficiently 
small numerical error) only the solutions suppressed on
the side where the initial data were specified. Therefore it is
necessary to repeat the search twice: for $v_I\approx v_R$ and
$v_I\approx -v_R$. 

\begin{figure}[tbh!]\begin{center}
  $a)$\hspace{8cm}$b)$
  \includegraphics[width=3.2in,angle=0]{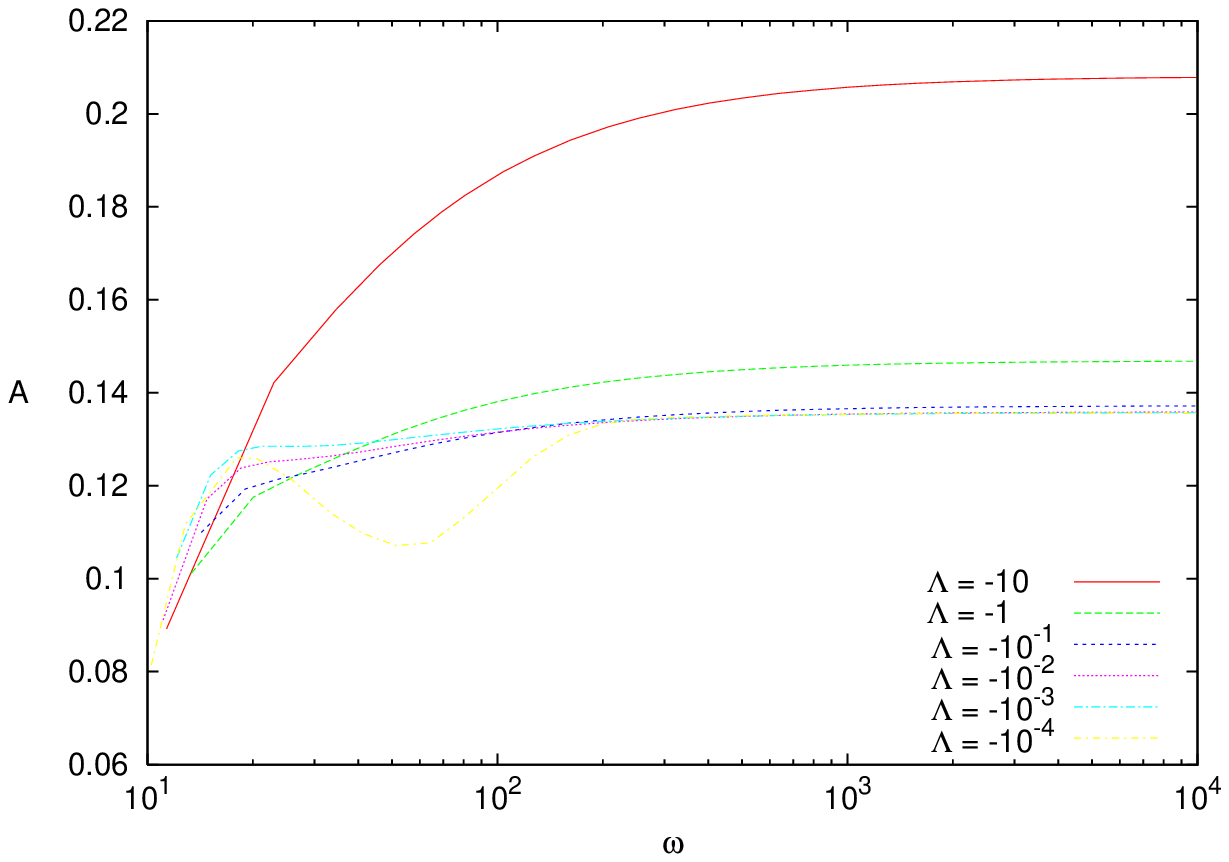}
  \includegraphics[width=3.2in,angle=0]{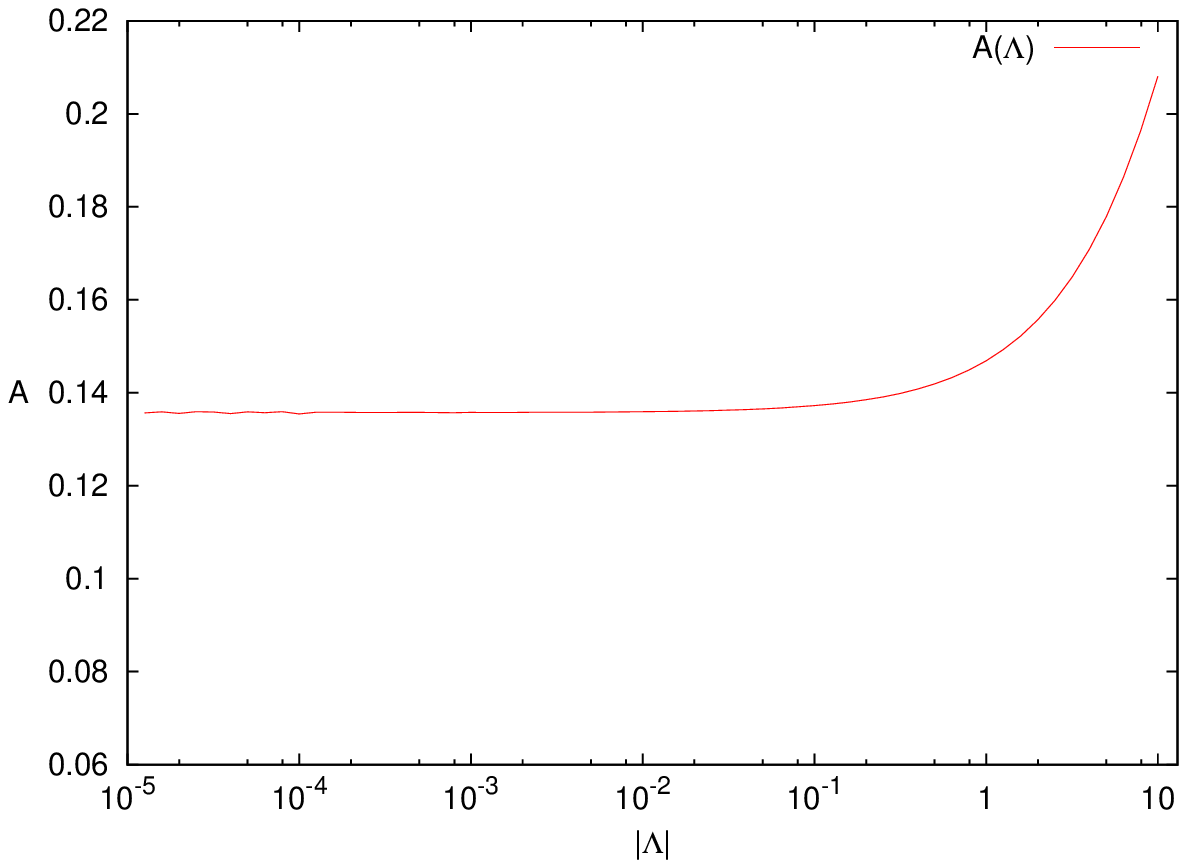}
  \caption{The rescaled eigenvalue separation correction term
    $A(\omega):=|\Delta\omega_n-\Delta\omega| 
    (\omega_n/\Delta\omega)^2$ is shown in $a)$ for several values of
    $\Lambda$. Its $\omega\to\infty$ limit is plotted in $b)$ as a
    function of $\Lambda$. The 'wiggles' at small values of
    $|\Lambda|$ are the results of numerical errors due to the precision
    limitation of the applied calculation method.}
  \label{fig:A}
\end{center}\end{figure}

The spectrum scan described above was performed for $18$
values of $\Lambda$ ranging from $-20$ to $-10^{-6}$. It revealed
the following properties (visualized in figs.~\ref{fig:eig-low} --
\ref{fig:A}).  
\begin{itemize}
  \item As analytically predicted, for each $\epsilon$ the spectrum of
    $\Theta$ is discrete and the eigenvalues are
    isolated. With the exception of the lowest $\omega$, the eigenfunctions are
    highly (exponentially in $\sqrt{\omega}$) suppressed for one triad
    orientation (sign of $v$). The eigenvalues corresponding to them
    are continuous functions of $\epsilon$. The density of eigenvalues
    is twice higher on the generic lattices than on the exceptional
    ones. Furthermore, for $\varepsilon=2$, the two families of left-suppressed
    and right-suppressed eigenfunctions converge (see fig.~\ref{fig:dw}a).  
  \item The separation $\Delta\omega_{n}:=\omega_{n+1}-\omega_n$ is
    not uniform. It depends on $\epsilon$ and $\Lambda$ as well as
    $n$. However, for large values of $\omega$, $\Delta\omega_n$
    converges to the limit value $\Delta\omega$ with convergence rate
    $\omega^{-2}$ (see fig.\ref{fig:A}a)
    \begin{equation}\label{eq:dw}
      \Delta\omega_n\ =\ \Delta\omega + O(\omega^{-2}) \ , \qquad
      \Delta\omega\ =\ \lim_{n\to\infty}\Delta\omega_n \ .
    \end{equation}
    Numerical inspection shows that the correction satisfies (with
    the exception of the lowest $\omega$) the following bound relation
    \begin{equation}\label{eq:A-bound}
      |\Delta\omega-\Delta\omega_n|\ \leq\ 
        \frac{A (\Delta\omega)^2}{\omega^{2}} \ ,
    \end{equation} 
    where, for $|\Lambda|<10$, $A<0.21$ and $A$ decreases for smaller
    $|\Lambda|$, reaching in the $|\Lambda|\to 0$ limit the value $A
    \approx 0.1358\pm 2\cdot 10^{-4}$ (see fig.\ref{fig:A}). 
  \item The limit $\Delta\omega$ was found numerically via $4$th order
    polynomial extrapolation of $\Delta\omega_n$. 
    It is a function of $\Lambda$
    only, i.e. it does not depend on $\epsilon$. Its values
    for different superselection sectors agree up to $10^{-9}$
    precision. The dependence on $\Lambda$ found numerically is shown
    on fig.~\ref{fig:dw}b. For small values of $|\Lambda|$ it can be
    approximated via a power function 
    \begin{equation}\label{eq:dwlow}
      \Delta\omega \approx a |\Lambda G|^b
    \end{equation}
    where $a\approx3.87$ and $b\approx0.0489$. 
\end{itemize}

The spectrum and normalizable eigenfunctions found here may be next
used to construct the semiclassical states. Details of this construction
will be presented in the next section.

\subsection{Semiclassical states, evolution}\label{sec:num-state}

Once we know the values of $\omega_n$ and $e_n(v)$, the construction of
a physical state from \eqref{eq:Psi} is straightforward. There are two
possibilities here: direct summation of equation 
(\ref{eq:Psi}b) or numerical integration via equation
(\ref{eq:Psi}a) (or equivalently via \eqref{eq:Theta}) of some initial data
specified at a given $\phi_o$. To find these data, we again have two 
methods at our disposal: one of them is the same direct summation of
(\ref{eq:Psi}b), but applied to one slice, whereas the second
possibility is the use of a slice of a WDW semiclassical state (see
section \ref{sec:WDW}) peaked at
large $v^{\star}$, where we do not expect strong quantum-geometric
effects. In practice we used the second method, integrating the state
in $\phi$ via equation \eqref{eq:Theta} and using as initial data both
the WDW slices and the results of the summation of (\ref{eq:Psi}b). The first
method of state calculation was used only to measure the wave packet
spread increase in large intervals of $\phi$, as the integration
methods were not precise enough for this application.

\subsubsection{Initial data}

Let us focus on the second method of initial data specification: constructing
the WDW slice. 
In order to be able to directly compare the dynamics of LQC model and
its WDW limit described in section \ref{sec:WDW} we take as the
initial data the $\phi=\phi_o$ section of the Gaussian state
\eqref{eq:WDWpsi} peaked at $p^{\star}_{\phi}=\hbar\omega^{\star}$ and
$v^{\star}$. Since \eqref{eq:Theta} is a second order equation, to
specify the initial data completely we also need $\dot{\Psi}(v,\phi_o)$
-- the first order derivative of $\Psi$ with respect to $\phi$. We get
it by integrating the integrand of \eqref{eq:WDWpsi},
multiplied by $i\omega$, over $\omega$.

In order to calculate the specified integrals, we first need to compute
the values of $\ub{e}_{\omega}(v)$, which are the normalized Bessel functions
$\BesK$ (see section \ref{sec:WDWsols}). To do so, we apply the
combined method specified by Gil, Segura and Temme \cite{BesK}.  

Once we have  $\ub{e}_{\omega}(v)$, we integrate \eqref{eq:WDWpsi} (and
the analogous expression for $\dot{\Psi}$) over the domain
$[\omega^{\star}-7\sigma, \omega^{\star}+7\sigma]$, using the trapezoid
method. Such choice of domain provides sufficient precision -- the errors
due to the removed tails are much smaller than the error associated
with the computation of the $\BesK$ functions. 

Note that we intend to construct the initial data corresponding to the
positive frequency solution to \eqref{eq:Theta}. In that case,
$\dot{\Psi}$ is already determined by $\Psi$ via (\ref{eq:Psi}a).
On the other hand, we determined it using positive frequency WDW equation
\eqref{eq:WDWmainpm}. Since $\sqrt{\Theta}$ differs from
$\sqrt{\ub{\Theta}}$, our initial data is not a pure positive frequency 
solution. To minimize the negative frequency part, we choose the following
method to construct states sharply peaked at large $v^{\star}$:
we require $v^{\star}$ be greater than $2.5p_{\phi}^{\star}/\hbar$, 
which keeps the negative frequency part below $10^{-3}$ of the
entire wave packet norm.  

We avoid the above problem if we use directly the basis of functions
$e_n(v)$ and sum them using (\ref{eq:Psi}b). In that case, as 
the spectral profile $\tilde{\Psi}_n$ we choose the restriction of the
Gaussian to $\{\omega_n\}$, that is
\begin{equation}\label{eq:tPsi}
  \tilde{\Psi}_n\ =\ e^{-\frac{(\omega_n-\omega^{\star})^2}{2\sigma^2}}
    e^{-i\omega_n\phi^{\star}}
\end{equation}
where $\hbar\omega^{\star}=p_{\phi}^{\star}$ is again the location of
the peak in the momentum and $\hbar\sigma/\sqrt{2}$ is its spread. 
The parameter $\phi^{\star}$ is determined by the position
  $v^{\star}$ of the peak in $v$ and value of $\phi_o$ via \eqref{eq:phistar}.
Similarly to the WDW initial data, we sum only over $\omega_n\in
[\omega^{\star}-7\sigma, \omega^{\star}+7\sigma]$. The derivative
$\dot{\Psi}$ is calculated by summing over the individual terms,
multiplied by $i\omega_n$. 

\subsubsection{Evolution}

Given some initial data, one can integrate it over some interval
$[\phi_o,\phi_1]$ using equation \eqref{eq:Theta}, which is a system of
a countable number of coupled ordinary differential equations (ODE). Due to
the $v$-reflection symmetry, it is enough to restrict the domain of
integration to $\lat^+=\lat_{+|v|}$ for generic lattices and
$\lat^+=\lat_{\epsilon}\cap\re^{+}$ for exceptional ones. 
Additional, the numerical nature of our study requires that we further
restrict the domain of $v$ to the finite subset $\lat^+_{v_{\max}} :=
\lat^+\cap[-v_{\max},v_{\max}]$, imposing at the outermost points of
the domain some (artificial) boundary conditions. Since the system
under consideration is a classically recollapsing one, it is enough to
choose the reflective conditions $\Psi = \dot{\Psi} = 0$. To prevent
their interference with the dynamics, we have chosen $v_{\max}$ to be
not smaller than $1.3v_R(\omega^\star)+2000$. 

\begin{figure}[tbh!]\begin{center}
  \includegraphics[width=3.2in,angle=0]{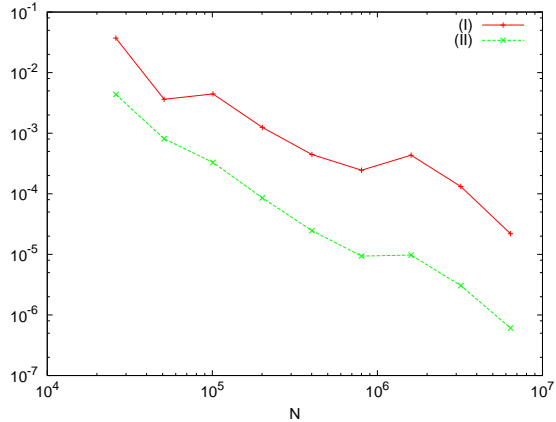}
  \caption{Convergence test for the integration method of a
    Gaussian wave packet generated with $\Lambda=-0.1$ and peaked at 
    $p^{\star}_{\phi}=10^3$, with relative $p_{\phi}$ spread $0.05$ and 
    $v^{\star}=0.5\,v_R(p^{\star}_{\phi})$. The initial data were
    specified at $\phi=0$ and evolved till $\phi=1$. The upper (red)
    curve shows the norm of the difference $\|\Psi_{(N)}-\Psi\|$ between
    the slice $\phi=1$ of the solution $\Psi_{(N)}$ corresponding to
    the integration with $N$ steps and the same slice of its $N\to\infty$
    limit $\Psi$ (found via $8$th order polynomial extrapolation). The lower
    (green) curve shows the analogous difference taken with respect to
    the metric \eqref{eq:norm2}.}
  \label{fig:conv}
\end{center}\end{figure}

Upon the above restriction of the $v$ domain, the equation
\eqref{eq:Theta} becomes a finite system of ODEs. We integrate it
using a $4th$-order adaptive Runge-Kutta method (RK4). To adapt the
steps of integration, we compare solutions corresponding to step
$\Delta\phi$ and $\Delta\phi/2$ and require the difference between
them (at a single $\Delta\phi$ step / two $\Delta\phi/2$ steps) to
satisfy the inequality:  
\begin{equation}
  \| \Psi_{\Delta\phi} - \Psi_{\Delta\phi/2} \|\ \leq\ 
  \frac{\epsilon\Delta\phi}{|\phi_1-\phi_o|} \|\Psi_{\Delta\phi/2}\|  
  \ ,
\end{equation}
where $\epsilon$ is a preset global bound. The two solutions are compared
via the following norm
\begin{equation}\label{eq:norm1}
  \| \Psi \|\ :=\ \sup_{v\in\lat^+_{v_{\max}}} |\Psi(v,\phi)| \ .
\end{equation}
Since only $|\Psi|$ enters the formulae for the expectation values of
$|\hat{v}|_{\phi}$ and $v^2_{\phi}$, it is also convenient to introduce
an auxiliary metric measuring the error in absolute values only
\begin{equation}\label{eq:norm2}
  \| \Psi_1 - \Psi_2 \|_{A}\ 
  =\ \sup_{v\in\lat^+_{v_{\max}}} \big||\Psi_1(v,\phi)|-|\Psi_2(v,\phi)|\big|
\end{equation}
An example of convergence test done with respect to both the norm
\eqref{eq:norm1} and the metric \eqref{eq:norm2} is shown in
fig.\ref{fig:conv}, where the results of integration with different
error bounds $\epsilon$ were compared against the result of polynomial
extrapolation at $\epsilon=0$. As we can see, the integration error is
for $|\Psi|$ at least one order of magnitude smaller than that for
$\Psi$ itself.

\subsubsection{Observables}\label{sec:lqc-obs}

Knowing an explicit form of $\Psi$ at
$\lat^+_{v_{\max}}\times[\phi_o,\phi_1]$, we can complete it to 
$(\lat^+_{v_{\max}}\cup\lat^-_{v_{\max}})\times[\phi_o,\phi_1]$ (where 
$\lat^-_{v_{\max}}:=\{-v:v\in\lat^+_{v_{\max}}\}$) via reflection and
find the expectation values of the observables \eqref{eq:obs}, restricting
the sums (\ref{eq:IP},~\ref{eq:expect}) to a finite domain
$\lat^+_{v_{\max}}\cup\lat^-_{v_{\max}}$. Their dispersions can be in
turn calculated in the standard way
\begin{subequations}\label{eq:disp}\begin{align}
  \expect{\Delta|\hat{v}|_{\phi}}^2\ &=\ \expect{\hat{v}^2_{\phi}}
  - \expect{|\hat{v}|_{\phi}}^2 &
  \expect{\Delta\hat{\phi}_{\phi}}^2\ &=\ \expect{\hat{\phi}^2_{\phi}}
  - \expect{\hat{\phi}_{\phi}}^2 \ .
\tag{\ref{eq:disp}}\end{align}\end{subequations}
where $\expect{\hat{v}^2_{\phi}}$, $\expect{\hat{\phi}^2_{\phi}}$ have
a form analogous to \eqref{eq:expect}.

In addition to $|\hat{v}|_{\phi}$, $\hat{\phi}_{\phi}$, it is useful to
introduce another family of observables: the regularized energy density
at a given moment of $\phi$
\begin{subequations}\label{eq:rho}\begin{align}
  \hat{\rho}_{\phi}\ 
  &:=\ \frac{1}{2\lPl^6}\left(\frac{6}{8\pi\gamma}\right)^3
  \hat{p}_{\phi}\,\hat{B}_{\phi}\, 
  \hat{p}_{\phi}\,\hat{B}_{\phi}
  \ , & 
  \hat{B}_{\phi}\Psi\ 
  &:=\ e^{i\sqrt{\Theta}(\phi-\phi')}B(v)\Psi(v,\phi') \ .
\tag{\ref{eq:rho}}\end{align}\end{subequations}
We calculate their expectation values via
\begin{equation}
  \sbra{\Psi}\hat{\rho}_{\phi}\sket{\Psi}\ =\ -
  \frac{K^2}{2\lPl^6}\left(\frac{6}{8\pi\gamma}\right)^3
  \sum_{\lat^+{v_{\max}}\cup\lat^-{v_{\max}}} B(v)
  |\Phi(v,\phi)|^2 \ ,\quad \Phi\ =\ 
  \partial_{\phi}(\widehat{|v|^{-1}}_{\phi}\Psi) \ ,
\end{equation}
whereas the dispersions can be derived analogously to \eqref{eq:disp}. 

The above methods for calculating the expectation values were applied to
the wave functions calculated earlier through the RK4 method. We analyzed
the states evolved (integrated) from both WDW and exact LQC Gaussian
wave packets corresponding to $17$ values of $\Lambda$ ranging
from $-20$ to $-10^{-6}$, for $5$ different superselection sectors
covering the full range of $\varepsilon$. The peak in momentum
$\phi^{\star}_{\phi}$ covered the values from $5\cdot 10^2$ to $10^4$ ($10$
values), while its relative spread ranged from $0.01$ to $0.1$.

\section{Results and discussion}\label{sec:concl}

\begin{figure}[tbh!]\begin{center}
  \includegraphics[width=6.0in,angle=0]{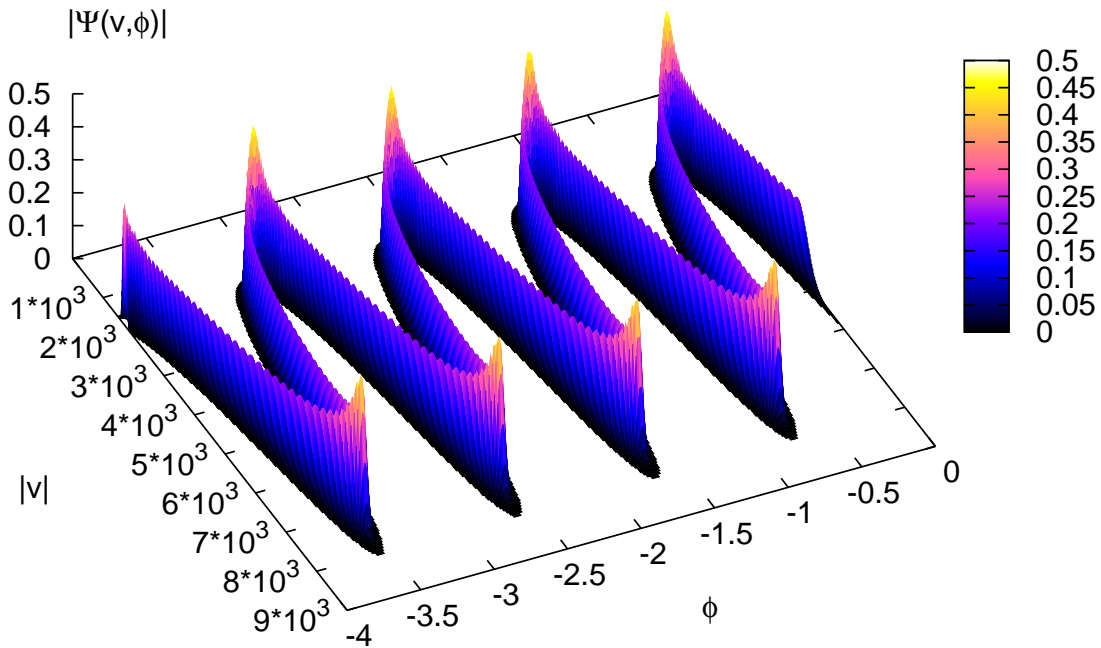}
  \caption{The absolute value of the wave function representing 
    a Gaussian state \eqref{eq:Psi} generated via backward integration
    of an initial profile corresponding to $\Lambda=-1$,
      $p_{\phi}^{\star}=5\cdot 10^3\hbar$, $\Delta 
      p_{\phi}/p_{\phi}^{\star}=0.01$, $v^{\star}=0.6\,v_R(p_{\phi}^{\star})$
    and evaluated at $\phi_o=0$. For presentation clarity, only values
    $>10^{-6}$ were shown on the plot.} 
  \label{fig:Psi-3d}
\end{center}\end{figure}

\begin{figure}[tbh!]\begin{center}
  $a)$\hspace{8cm}$b)$
  \includegraphics[width=3.2in,angle=0]{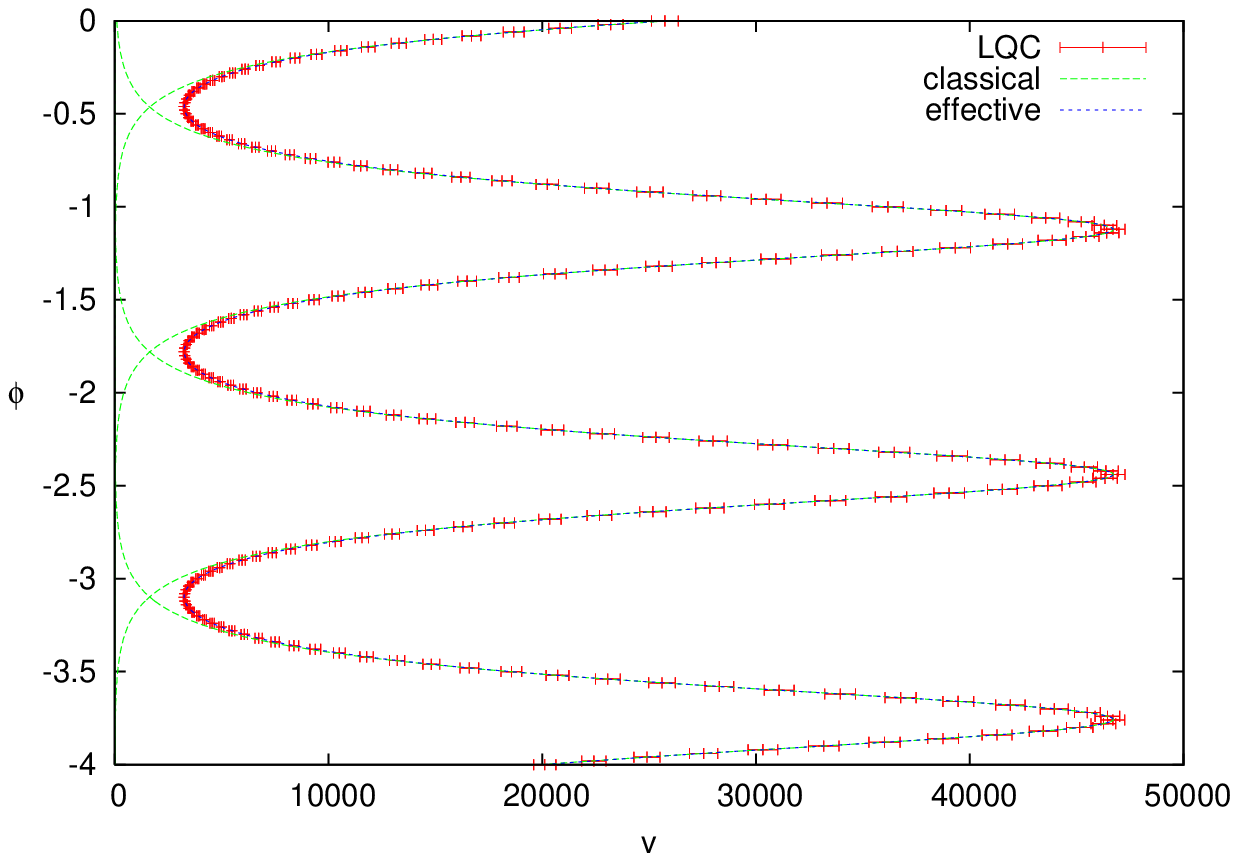}
  \includegraphics[width=3.2in,angle=0]{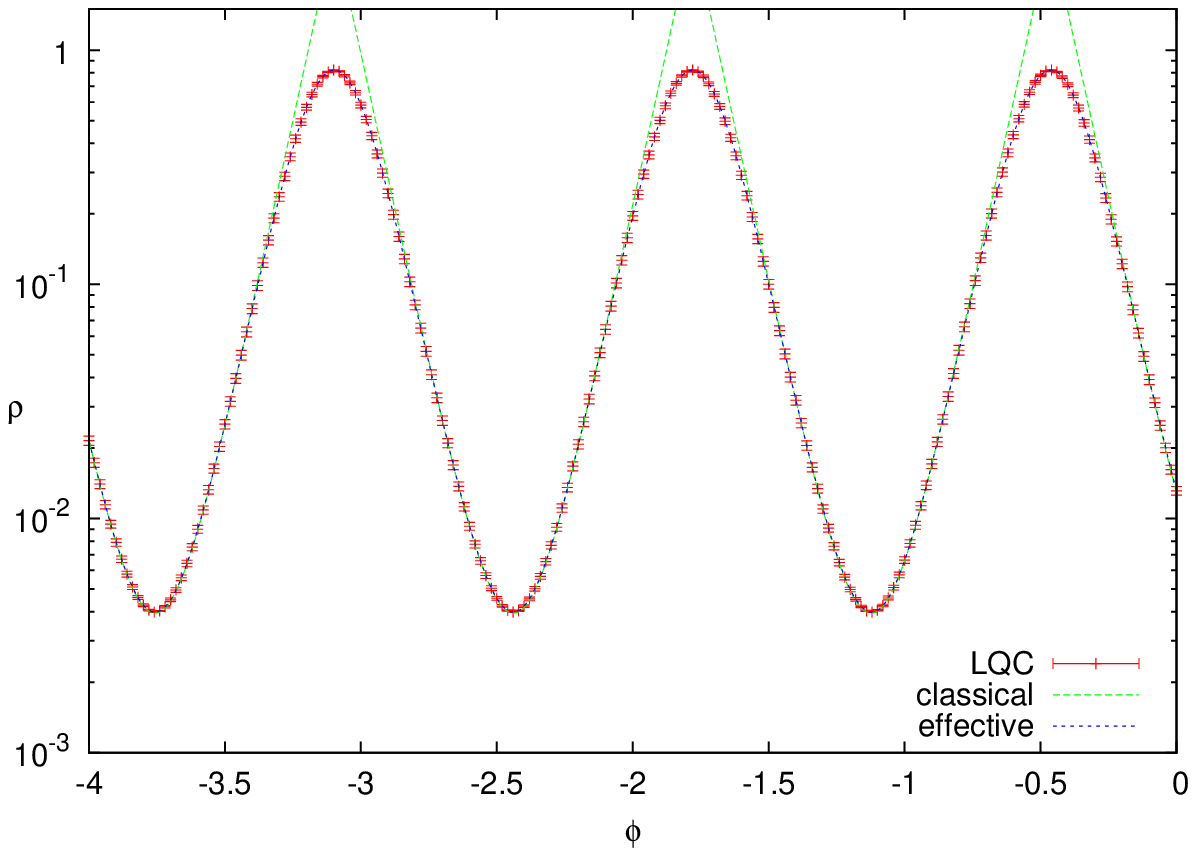}
  \caption{The expectation values (red bars) of $|\hat{v}|_{\phi}$ $(a)$ and
    $\hat{\rho}_{\phi}$ $(b)$ are compared against
    the classical (red lines) and effective (blue lines) trajectories of
    $v(\phi)$ and $\rho_{\phi}(v)$ respectively. The data corresponds
    to a Gaussian wave packet \eqref{eq:Psi} with $\Lambda=-0.1$, 
    $p^{\star}_{\phi}=10^4\hbar$, $\Delta p_{\phi}/p^{\star}_{\phi}=0.012$, 
    $v^{\star}=0.55\,v_R(p^{\star}_{\phi})$ 
    specified at $\phi_o = 0$ and evolved backwards. Because of the large
    changes in magnitude of $\expect{\rho_{\phi}}$, a logarythmic
    scale was used for the $y$-axis of fig.$b)$.
  }
  \label{fig:vr-traj}
\end{center}\end{figure}

\begin{figure}[tbh!]\begin{center}
  $a)$\hspace{8cm}$b)$
  \includegraphics[width=3.2in,angle=0]{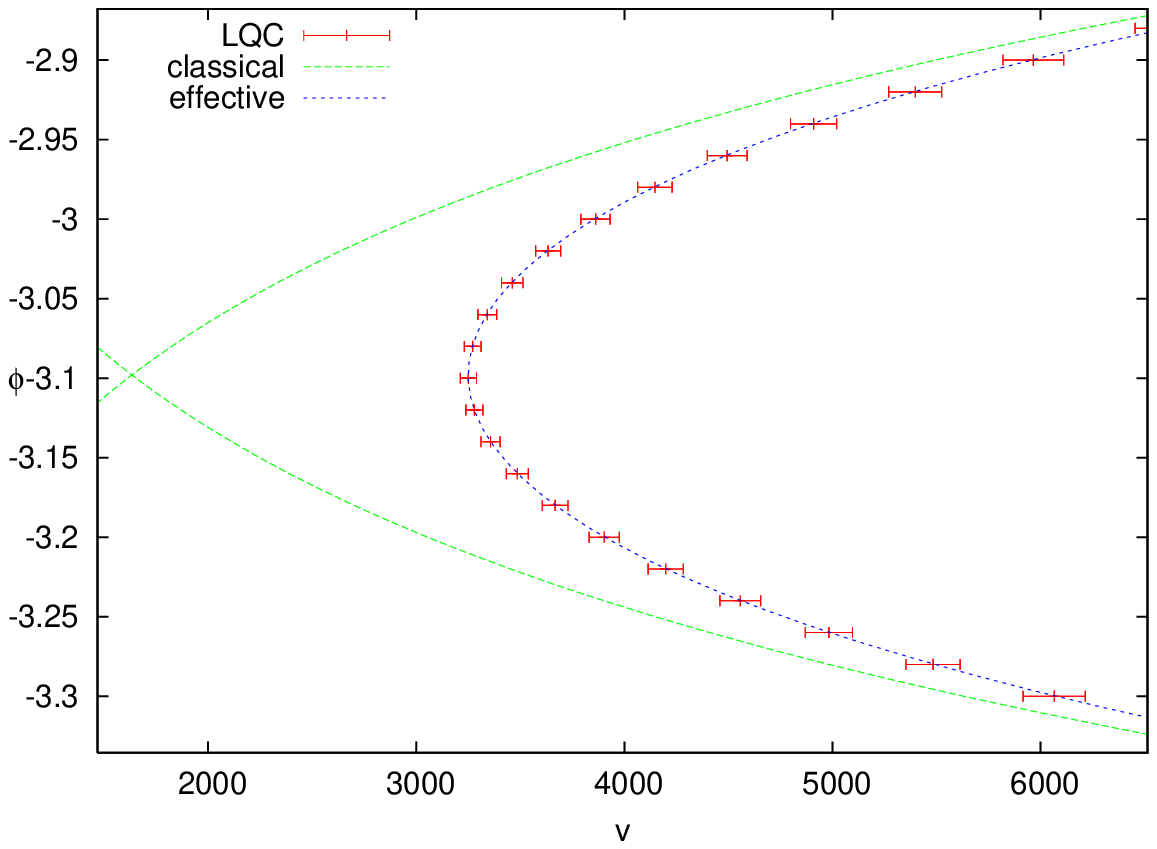}
  \includegraphics[width=3.2in,angle=0]{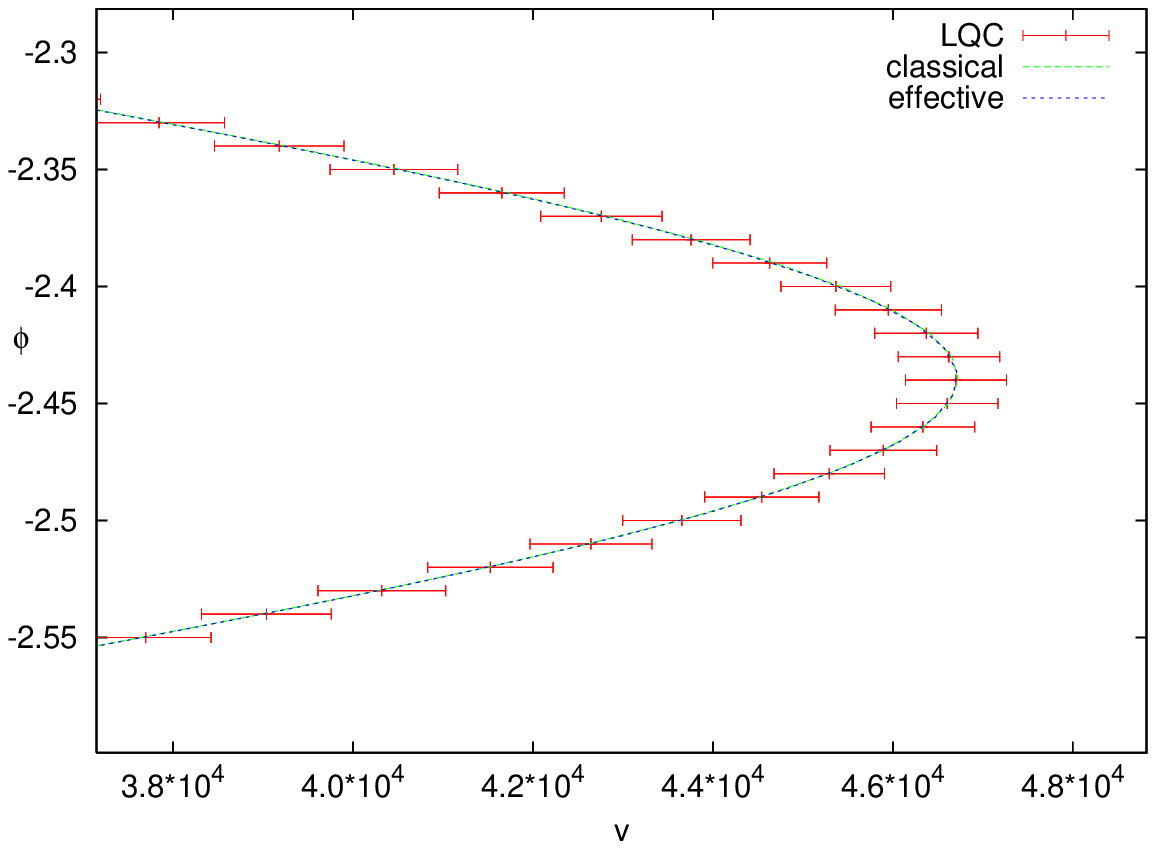}
  \caption{A detailed picture of the comparizon of
    $\expect{|v|_{\phi}}$ against the $v(\phi)$ trajectories presented in
    fig.~\ref{fig:vr-traj}a is shown near the bounce $(a)$ and recollapse
    $(b)$ points respectively.} 
  \label{fig:v-zoom}
\end{center}\end{figure}

\begin{figure}[tbh!]\begin{center}
  $a)$\hspace{8cm}$b)$
  \includegraphics[width=3.2in,angle=0]{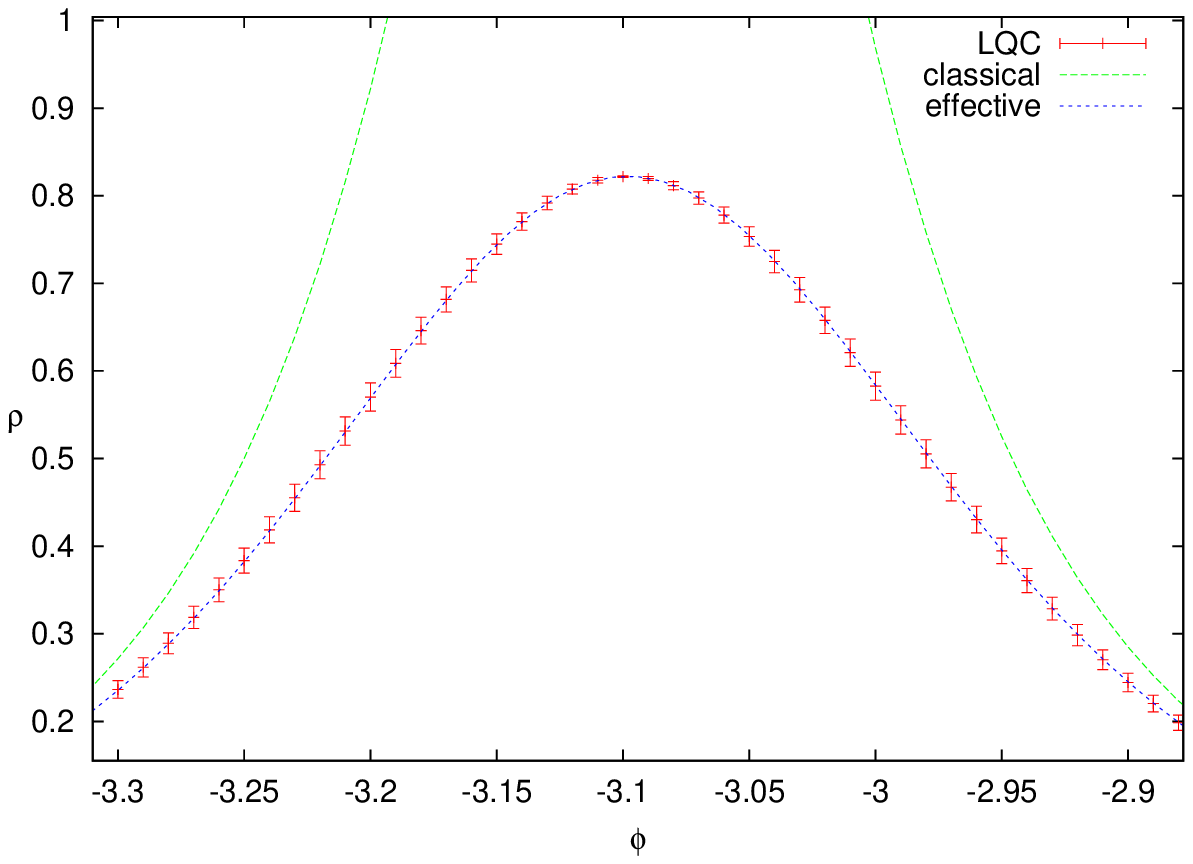}
  \includegraphics[width=3.2in,angle=0]{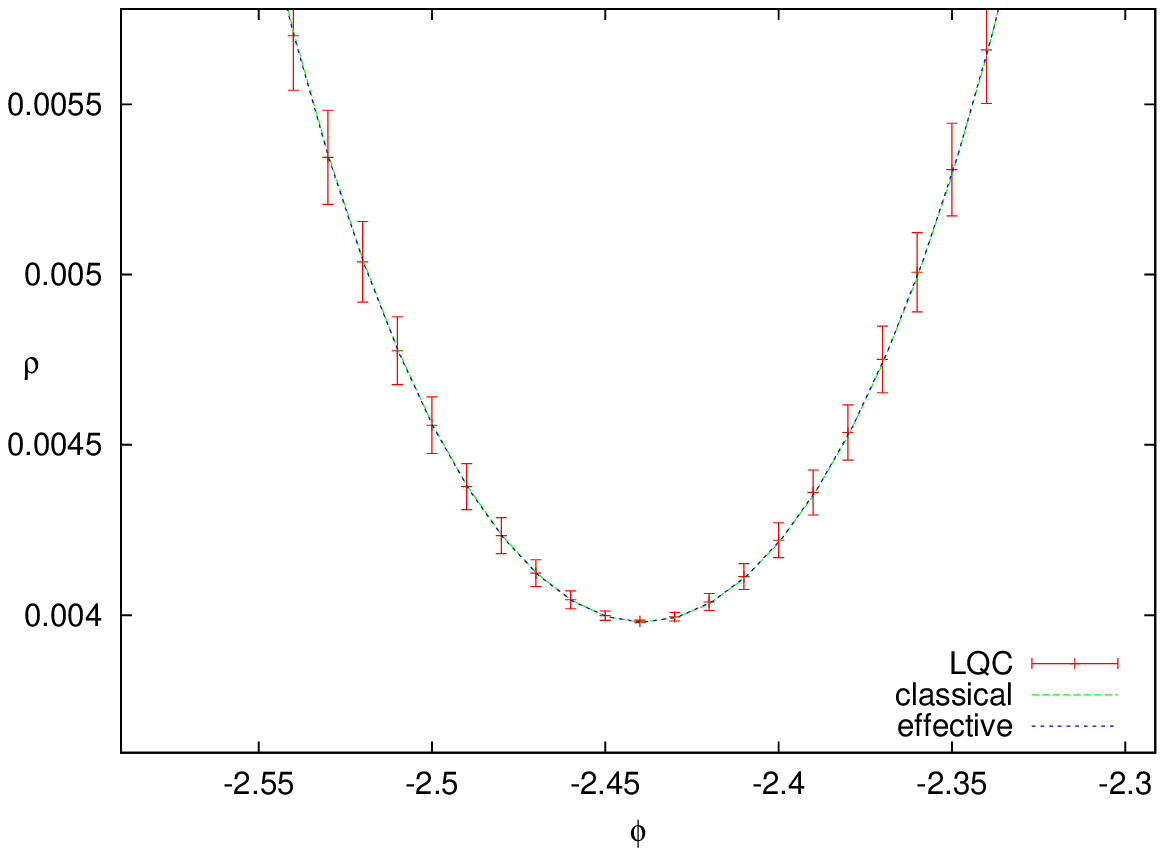}
  \caption{A detailed picture of the comparizon of
    $\expect{\rho_{\phi}}$ against the $\rho_{\phi}(\phi)$ trajectories
    presented in fig.~\ref{fig:vr-traj}b is shown near the bounce $(a)$ and
    recollapse $(b)$ points respectively.}
  \label{fig:rho-zoom}
\end{center}\end{figure}

An example of the results of our numerical investigations is presented
in figs. \ref{fig:Psi-3d} -- \ref{fig:rho-zoom}.
The general properties of the considered model are similar to the ones of
the models previously investigated: $\Lambda=0$ and $k=1$, that is:
\begin{itemize}
  \item The states remain sharply peaked for long
    evolution times. On each superselection sector and large $\omega$, the
    spectrum of $\Theta$ quickly approaches uniformity (with approach
    rate $\omega^{-2}$). In consequence, a wave packet sharply peaked
    at large $p_{\phi}$ should be almost periodic in $\phi$. This
    expectation is confirmed by our numerical results, where already for
    $p^{\star}_{\phi}$ of the order of few thousands the departures
    from periodicity were undetectable within given precision
    of integration. 
  \item For large volumes (small energy densities), the trajectory of the
    expectation values $\expect{|v|_{\phi}}$ agrees with the classical
    one given by \eqref{eq:WDWtraj}. In particular, the universe
    recollapses at the volume predicted by the classical theory even
    for large values of $\Lambda$; this was numerically confirmed
    for $|\Lambda|$ up to $20$. 
  \item Once the expectation value of the energy density approaches
    the Planck order, we observe the departures from the classical
    theory due to quantum-geometric corrections. The corrections act
    effectively like an additional repulsive force, which in
    particular causes the bounce at the point where the total energy
    density $\expect{\hat{\rho}_{\phi}} + \Lambda/(8\pi G)$ approaches
    a critical value $\rho_c\approx 0.82\rho_{\Pl}$, identified
    already in \cite{aps-imp}. 
  \item After the bounce, the universe again enters (another) classical
    trajectory repeating the cycle of expansion, recollapse and
    contraction till the energy density grows again to Planck scale. In
    consequence, the evolution is periodic and, similarly to the $k=1$ 
    case, we are dealing with a cyclic model.
  \item The wave packet remains sharply peaked even in the region
    where the quantum corrections are strong. In consequence, the evolution
    can be described by the classical effective dynamics, similarly to the
    $\Lambda=0$ case. Indeed, the comparison of the values of
    $\expect{|v|_{\phi}}$ with the effective trajectories corresponding to
    the holonomy corrections (see Appendix \ref{sec:eff}) has
    shown that they agree up to an error well below $\expect{\Delta v}$.
\end{itemize}

The results listed above show that the picture based on the analysis of the
previous models is valid here as well. Similarly to that cases, the
correction due to the discreteness of geometry cause gravity to become 
repulsive at large energy densities and, in particular, force the
universe to bounce when the energy density reaches a critical value. This
indicates that $\rho_c$ may be a fundamental quantity, independent on
the matter content at least in isotropic cases. Furthermore, the
states remain sharply peaked even in regions where quantum-geometric
effects dominate the dynamics, where in principle one expects to
loose the semiclassicality. The dynamics itself can be well approximated by 
an effective Friedmann equation (see Appendix \ref{sec:eff})
\begin{equation}
  H^2 = \fracs{8\pi G}{3}\rho(1-\rho/\rho_c) \ ,
\end{equation}
where $H$ is a Hubble rate and $\rho$ is a total energy density.

The agreement between the quantum evolution and the effective one brings out
another issue: since the spectrum of $\Theta$ is not exactly uniform, the
states are not exactly periodic and a spread increase can be observed
between cycles. This leads ultimately to the loss of semiclassicality.
This in turn raises the question about the size of time interval in which
the state remains sharply peaked.

To answer this question, we analyzed the spread increase within one cycle of
evolution. It can be estimated via the heuristic methods
described in section \ref{sec:heuDisp} and turns out to be much smaller than in
$k=1$ case. For example, when $\Lambda \approx -10^{-120}$, a
universe peaked about $p^{\star}_{\phi}$ large enough for it to grow
to megaparsec size, and with relative dispersions in $p_{\phi}$ and
$v$ of the same order, will need at least $10^{70}$ cycles for the
relative dispersion to double. The number of cycles needed to grow
to a considerably large value (say $10^{-6}$) is correspondingly larger. 

For small values of the momentum (that is $p_{\phi}^{\star}
\leq 10^3\hbar$), we were able to confirm the heuristics numerically. 
Also, since for larger momenta the states become more and more
semiclassical, we expect the estimate to become more accurate 
there. The result is however far from general, as numerical tests were 
done for a specific family of states only, thus (as it was discussed
in \cite{mb-L}) do not allow us to exclude the situation, where some
specific example of state violates the bound. On the other hand, the
proposed estimate is based on the properties of the spectrum of
$\Theta$, thus we expect that a bound of at least a similar order
should hold in general. Such situation happened for example in the
$\Lambda=0$ case \cite{cs-rec}, where it was possible to find (in the
context of sLQC) an analogous bound satisfied by all the states
which admit semiclassical epoch (see section \ref{sec:intro}) in their
history. A similar bound was next derived in exact LQC
\cite{kp-rec}. More precise statements regarding model considered here
will however require further work.\\  

\noindent{{\bf Acknowledgments}} We would like to thank Abhay Ashtekar,
Wojciech Kami\'nski and Parampreet Singh for extensive discussions and
helpful comments. We also profitted from discussions with Martin
Bojowald and Jerzy Lewandowski. This work was supported in part by the
National Science Foundation (NSF) grant PHY-0456913 and the Eberly
research funds. TP acknowledges financial aid provided by the I3P
framework of CSIC and the European Social Fund. EB acknowledges the
support of the Center for Gravitational Wave Physics, funded by the
National Science Foundation under Cooperative Agreement PHY-0114375.

\appendix

\section{Anti-symmetric sector of LQC}\label{sec:antisymm}

Due to the lack of a parity violating interaction in the model considered in
this article, the change in the triad orientation is a large gauge symmetry. 
This allowed us to restrict the physical Hilbert space to the subspace
of states invariant with respect to the reflection in $v$
corresponding to this orientation change -- the symmetric sector. In
principle, however, we could choose instead the space of states which
are antisymmetric under considered transformation. There is no
physical reason to favor one of these two choices over the other. This
raised a concerns on whether the results of LQC are tied to the
selection of symmetric sector and whether they will still hold in the 
antisymmetric one. We address these concerns here by repeating the
constructions of section \ref{sec:LQC}, this time building the
physical Hilbert space out of antisymmetric states. 

First, following section \ref{sec:LQC} we divide the kinematical
Hilbert space $\Hilkg$ onto superselection sectors, i.e.~the spaces $\Hilkge$
of functions supported on lattices $\lat_{\varepsilon}$. The results
of \cite{klp-nL} (self-adjointness of $\Theta$ and discreteness of its
spectrum on each of these spaces) were derived without any symmetry
assumption, thus they hold also in our case. Furthermore, as we will show
below, the spectrum is non-degenerate also when we restrict the space
of eigenfunctions to the antisymmetric ones. In consequence, we can
construct the physical Hilbert space as specified in \eqref{eq:Psi},
but by imposing on the relevant eigenfunctions $e_n^a$ the condition
$e^a_{n}(v) = -e^a_{n}(-v)$ instead of the symmetry requirement. 

\begin{figure}[tbh!]\begin{center}
  $a)$\hspace{8cm}$b)$
  \includegraphics[width=3.2in,angle=0]{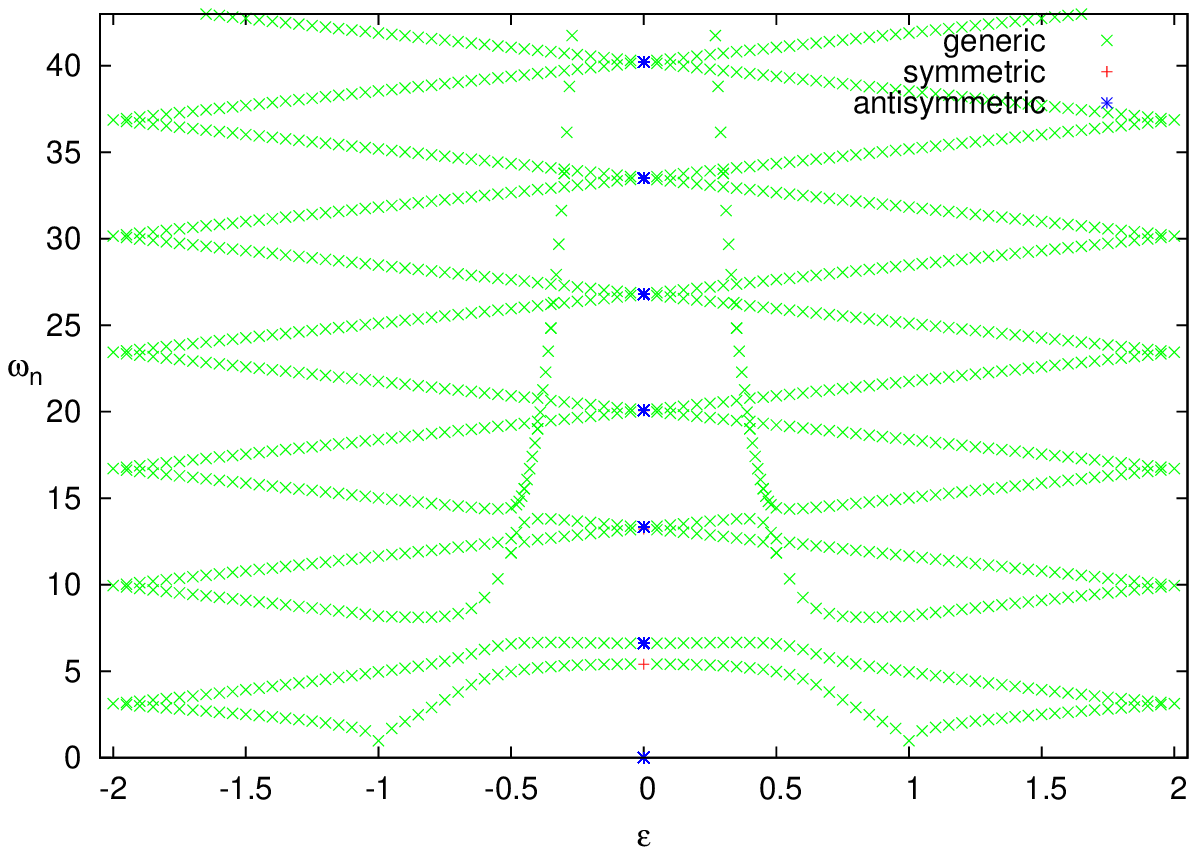}
  \includegraphics[width=3.2in,angle=0]{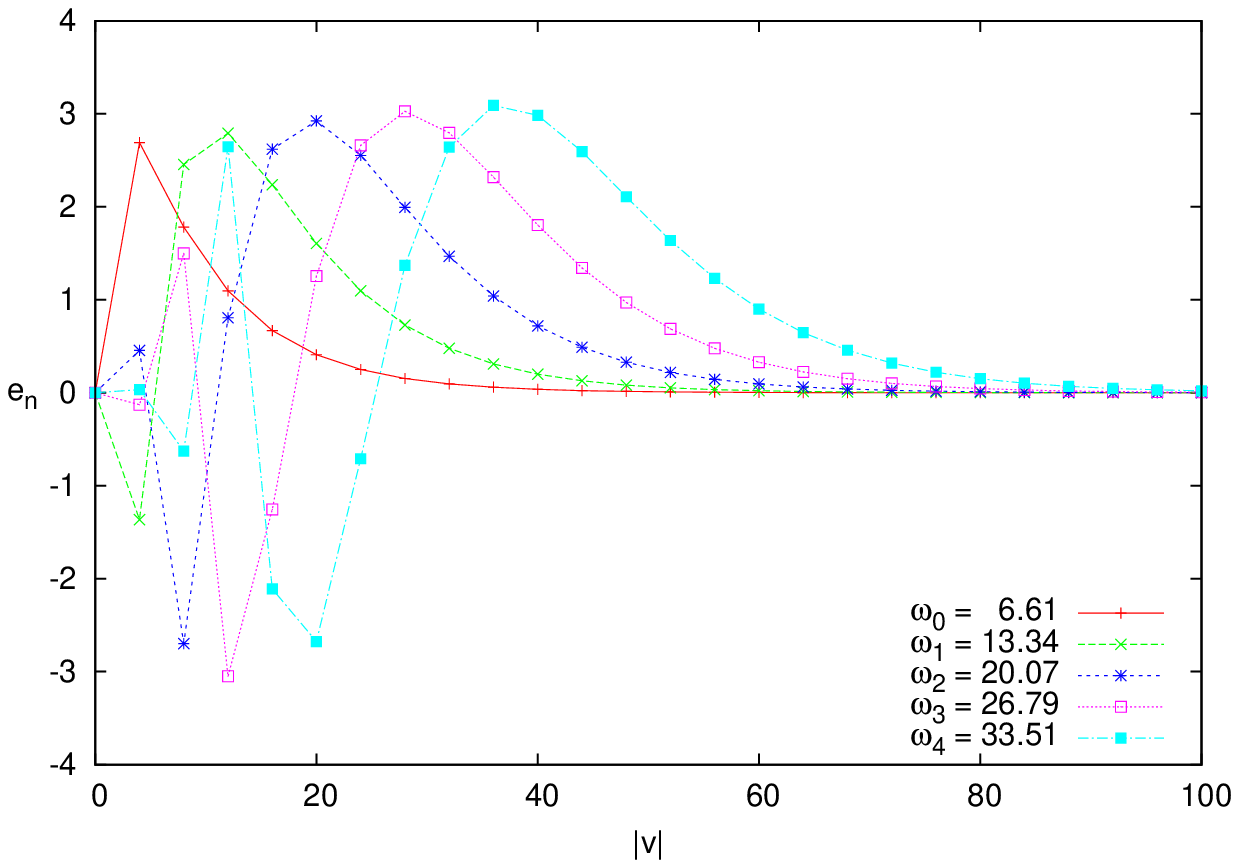}
  \caption{$a)$ A set of lowest ($\omega<44$) spectrum elements
    of $\Theta$ for the symmetric and antisymmetric sector (with
    $\Lambda=-1$) is 
    shown with respect to $\pm|\varepsilon|$. The green x-es 
    represent the eigenvalues corresponding to the cases where the relation
    between symmetric and antisymmetric eigenfunctions is given by
    (\ref{eq:sa-eg}, \ref{eq:sa-e2}) (denoted as {\it generic}). The
    red crosses and blue stars represent the eigenvalues of, respectively,
    the symmetric and antisymmetric eigenfunctions on the lattice
    $\lat_{\varepsilon=0}$. The 
    antisymmetric eigenfunctions $e^a_{0}$ to $e^a_{4}$ corresponding
    to the eigenvalues shown in $a)$ are presented in $b)$. For clarity,
    they are plotted on the $v>0$ semiaxis only.} 
  \label{fig:antisymm}
\end{center}\end{figure}

To check the effect of the above modification on the dynamics, we have to
examine how it changes the exact form of $e_n$. That, in turn, depends
on the value of the superselection sector label $\varepsilon$.
\begin{itemize}
  \item For $\epsilon\neq 0,2$ (generic lattices), the symmetric
    eigenfunction on $\lat_{\varepsilon}$ is completely determined (see
    discussion in section \ref{sec:num-eig}) by its restriction to the
    lattice $\lat_{+|\varepsilon|}$, with the remaining part supported
    on $\lat_{-|\varepsilon|}$ determined via a symmetry
    relation. Furthermore, symmetry does not impose any constraint on
    the part supported on $\lat_{+|\varepsilon|}$ itself and we can
    complete it to the antisymmetric eigenfunction by simply acting with
    $-\Pi$ on it. In consequence, there exists a $1-1$ correspondence between
    these two types of eigenfunctions. Namely, each antisymmetric
    eigenfunction $\psi^a$ is related to the symmetric one $\psi$ via: 
    \begin{equation}\label{eq:sa-eg}
      \psi^a(v)\ =\ \begin{cases} \psi(v) & v\in\lat_{+|\varepsilon|} \\
      -\psi(v) & v\in\lat_{-|\varepsilon|} \end{cases} \ ,
    \end{equation}
    This implies that, in the antisymmetric sector, the spectrum of
    $\Theta$ is the same as in the symmetric one. 
  \item When $\epsilon=2$, the situation is similar to the generic case. 
    The solutions to \eqref{eq:eigen} on two sublattices
    $\lat_{\varepsilon}\cap\re^+$ and $\lat_{\varepsilon}\cap\re^-$
    are independent, thus each eigenfunction is again determined by
    its restriction to $\lat_{\varepsilon}\cap\re^+$. In consequence,
    we again have the $1-1$ correspondence between the symmetric and
    antisymmetric eigenfunctions
    \begin{equation}\label{eq:sa-e2}
      \psi^a(v)\ =\ \begin{cases} \psi(v) & v>0 \\
      -\psi(v) & v<0 \end{cases} \ ,
    \end{equation}
    and the spectra of $\Theta$ in both sectors are identical.
  \item The case when $\epsilon=0$ requires a bit more care. In section
    \ref{sec:num-eig}, the symmetry assumption imposed on solutions to
    \eqref{eq:eigen} an additional constraint, allowing to determine
    $\psi(4)$ for known $\psi(0)$. Antisymmetry replaces this
    constraint by a different one: $\psi^a(0)=0$. Therefore the whole
    procedure of identifying the normalizable eigenfunctions has to be
    redone. We can however apply exactly the same procedure as in
    section \ref{sec:num-eig}. The results are as follows:
    \begin{itemize}
      \item The qualitative features of the eigenfunctions remain the
        same. In particular, we can still distinguish the same 5 zones of
        exponential/oscillatory behavior (see fig.\ref{fig:antisymm}a). 
        Their boundaries $v_B$, $v_R$ are exactly the same as in the 
        symmetric case.
      \item The spectrum of $\Theta$ in the antisymmetric sector is
        different than in the symmetric case, however the eigenvalues of
        one sector approach the ones of the other very quickly (see
        fig.\ref{fig:antisymm}b). In consequence, the separation
        between the eigenvalues approaches, as $\omega\to\infty$, the same
        limit shown in fig.~\ref{fig:dw}b. The rate of approach to
        this limit also remains the same.
    \end{itemize}
\end{itemize}

The similarity between eigenfunctions of the two considered sectors
implies an analogous similarity between the physical states. In particular,
for $\varepsilon\neq 0$, if the eigenfunctions satisfy equations 
(\ref{eq:sa-eg}-\ref{eq:sa-e2}), so will the wave
functions. Then if we take two physical states, a symmetric and
an antisymmetric one with the same spectral profile $\tilde{\Psi}_n$, the
expectation values of (all the powers of) the observables defined in
section \ref{sec:lqc-obs} will be {\it exactly the same} for both of
them. 

For $\varepsilon=0$, due to the slight difference in the spectrum, we have to 
repeat the analysis of \ref{sec:num-state}. But again the numerical
checks reveal no measurable deviations from the results obtained in the symmetric
sector. 

In summary the results obtained for both $\varepsilon\neq 0$ and
$\varepsilon=0$ show that working  with the antisymmetric sector instead of
the symmetric one does not produce any qualitative changes or (apart from
a slightly different spectrum of $\Theta$ in $\varepsilon=0$ case)
any measurable modifications to the physics predicted by the model.

\section{Heuristic description}\label{sec:heu}

In this sections we discuss some issues related to the heuristic method
for the description of the quantum evolution. We divide its content
into two parts, dedicated respectvely to the effective classical
dynamics and the estimate of the dispersion growth during the
evolution of the semiclassical state.

\subsection{Effective dynamics}\label{sec:eff}

The numerical tests described in the main body of the paper have shown that
if a state is semiclassical at some epoch, it will remains so for 
a large fraction of the evolution (i.e., a large number of cycles of
bounces and recollapses). 
In particular it remains sharply peaked even in the regions where the quantum
gravity corrections modify the dynamics. This indicates the existence
of a (n effective) classical theory whose predictions well agree with
those of LQC.

At the rigorous level, such theory was derived for $\Lambda=0$
\cite{VT} with the use of the geometric formulation of quantum mechanics
\cite{GQM}. Up to the second order quantum corrections (remaining 
always small during the evolution), its results confirm the
predictions of the classical effective dynamics proposed earlier 
\cite{pv,aps-imp}, derived heuristically by replacing the connection $c$ in
classical Hamiltonian by $\sin(\bar{\mu}c)/\bar{\mu}$. Here we apply
this heuristic method to the system with a cosmological constant
considered in the main body of the paper. An analogous derivation of
the effective dynamics (to the level of quadratures) and the analysis of
the resulting trajectories was done in \cite{eff}, however the
trajectory parametrization used there makes the direct comparizon with
the results of quantum evolution difficult.  

The classical Hamiltonian of the system is related to the constraint
(\ref{eq:classHgrav}, \ref{eq:classHcompl}) via $\Heff = C/(16\pi G)$. 
Application of the rule $c \to \sin(\bar{\mu}c)/\bar{\mu}$ yields the
result 
\begin{equation}\label{eq:Heff}
  \Heff\ =\ 
  -\frac{3}{8\pi G\gamma^2\bar{\mu}^2}|p|^{\frac{1}{2}}\sin^2(\bar{\mu}c) 
  + \frac{1}{2}\frac{p^{2}_{\phi}}{|p|^{\frac{3}{2}}}
  + \frac{p^{\frac{3}{2}}}{16\pi G} \Lambda \ .
\end{equation}
Hamilton's equations $\dot{v} = \{v,\Heff\}$ and $\dot{\phi} =
\{\phi,\Heff\}$ are identical to the $\Lambda=0$ case. Written in terms
of $v$ they are respectively
\begin{subequations}\label{eq:Ham}\begin{align}
    \label{eq:vdot}
  \dot{v}\ &=\ \frac{3v}{\sqrt{2\sqrt{3}\pi\gamma\lPl^2}} 
    \sin(\bar{\mu}c) \cos(\bar{\mu}c) \ , &
  \dot{\phi}\ &=\ \left(\frac{6}{8\pi\gamma\lPl^2}\right)^{\frac{3}{2}}
    \frac{K}{|v|}p_{\phi} \ . 
\tag{\ref{eq:Ham}}\end{align}\end{subequations}  
Taking the square of (\ref{eq:Ham}a) and supplying $\sin(\bar{\mu}c)$
via \eqref{eq:Heff}, we arrive to an analog of Friedmann equation:
\begin{equation}\label{eq:FriedEff}
  H^2\ :=\ \left(\frac{\dot{v}}{3v}\right)^2\ =\
  \frac{8\pi G}{3} \rho \left( 1-\frac{\rho}{\rho_c} \right) \ , 
\end{equation}
where $\rho$ and $\rho_c$ are the total matter energy density and the critical
energy density found in \cite{aps-imp}
\begin{subequations}\label{eq:RhoRho}\begin{align}
  \rho\ &:=\ \rho_{\phi}+\frac{\Lambda}{8\pi G} \ , &
  \rho_{\phi}\ &:=\ \frac{p^2_{\phi}}{2p^3} \ , &
  \rho_c\ &:=\ \frac{\sqrt{3}}{16\pi^2\gamma^3 G^2\hbar} \ .
\tag{\ref{eq:RhoRho}}\end{align}\end{subequations}
Applying (\ref{eq:Ham}b), we can rewrite the resulting Friedmann
equation in the form involving $v$ and $\phi$ only
\begin{equation}\label{eq:Fried1}
  v_{\phi}\ =\ \pm\, v\,\sqrt{12\pi G} \left[ \frac{\rho}{\rho_{\phi}}
  \left(1-\frac{\rho}{\rho_c}\right)\right]^{\frac{1}{2}} \ .
\end{equation}
The sign in front of the right hand side of the above equation depends on
the evolution epoch, and in particular changes during
recollapse. Therefore, it is convenient to rewrite \eqref{eq:Fried1} in
the second order form (obtained by differentiating it):
\begin{equation}\label{eq:Fried2}
  v_{\phi\phi}\ =\ 12\pi G\, v 
  \left[ \left(\frac{2\rho}{\rho_{\phi}}-1\right)
         \left(1-\frac{\rho}{\rho_c}\right) 
         + \frac{\rho}{\rho_c} \right] \ .
\end{equation}

To compare the results of the numerical evolution, we integrated
equation \eqref{eq:Fried2} numerically using a fifth-order adaptive
Runge-Kutta method (known as {\it RK45}). The initial value $\dot{v}$,
needed to complete the initial data specification, was calculated via
\eqref{eq:Fried1}. 

An example of the comparison results is shown in
figs.~\ref{fig:vr-traj},~\ref{fig:v-zoom},~\ref{fig:rho-zoom}. The 
trajectories agree with the results of the quantum evolution everywhere. 
The differences between them are much smaller than the spreads of the wave
packets even near the bounce.

\subsection{Bound on the dispersion growth}\label{sec:heuDisp}

The analysis of section \ref{sec:num} has shown that the states that are
semiclassical at a given initial time $\phi_o$ remain so for many cycles
of bounces and recollapses. However, due to non-uniformity of the
spectrum of the $\Theta$ operator, the initially coherent wave packet slowly
spreads out. Here we derive an upper bound on this spread growth using
some heuristic estimates and applying the knowledge about the spectrum
of $\Theta$ presented in section \ref{sec:num-eig}.

To start with, let us note that for large $p_{\phi}=\hbar\omega$ the
distance between neighboring eigenvalues is almost constant
$\omega_{n+1}-\omega_n \approx \Delta\omega$ (see \eqref{eq:dw}). In
consequence, the wave function is almost periodic in $\phi$, with an
approximate period equal to
\begin{equation}\label{eq:tw}
  T\ \approx\ \frac{2\pi}{\Delta\omega} \ .
\end{equation}
Now, if we consider two classical (effective) trajectories
corresponding to $p_{\phi}$ equal respectively to $\hbar\omega$ and
$\hbar\omega+\delta\omega$, the difference between periods is
determined by the corrections to the uniformity of
$\Delta\omega_n$. They are in turn bounded by the function
$A\,(\Delta\omega)^2\,\omega^{-2}$ (see \eqref{eq:A-bound}). Applying
this bound to \eqref{eq:tw} (i.e. taking $\Delta\omega_n =
\Delta\omega(1+A\omega^{-2})$) and neglecting terms of higher order in
$\delta\omega$, we obtain the following difference in $T$:  
\begin{equation}
  \delta T\ \approx\ \frac{4\pi A}{\omega^3}\delta\omega \ ,
\end{equation}
which can be now used to estimate the growth of $\Delta v/v$ within one
cycle. To do so, we note that, since the cosmological constant term acts
like a positive $v^2$ potential, the speed $v_{\phi}$ is bounded from
above by the speed $v^o_{\phi}$ of a classical universe with $\Lambda=0$
\begin{equation}
  |v_{\phi}|\ \leq\ |v^o_{\phi}|\ :=\ \sqrt{12\pi G}|v| \ .
\end{equation} 
In consequence:  
\begin{equation}\label{eq:ddv-bound}
  \frac{\delta v}{v}\ \leq\ 8\sqrt{3}\pi^{\frac{3}{2}}A\,
             \frac{1}{\omega^2}\,
             \frac{\delta\omega}{\omega} \ .
\end{equation}
  
In order to arrive to this bound, we used some heuristic methods that need to
be confirmed using numerics. Unfortunately, due to the extremely small
values of $\delta v/v$, we were able to check \eqref{eq:ddv-bound} only
for small values of the frequency, $\omega\leq 10^3$. To do so, we
calculated the semiclassical states in two intervals of $\phi$
separated by a large ($>100$) distance in $\phi$ and compared the
difference between the maximal relative dispersion in $v$ observed 
within one cycle in both of the chosen intervals. To compute the wave
functions, we used a direct summation method specified in section
\ref{sec:num-state}. Within the checked range $500\leq\omega\leq 1000$, the
bound was satisfied.


\begin{thebibliography}{99}
  \bibitem{lqc} M.~Bojowald, {Loop Quantum Cosmology}, Living
    Rev.Rel. \textbf{8}, 11 (2005), \texttt{arXiv: gr-qc/0601085};\\
    A.~Ashtekar, {An Introduction to Loop Quantum Gravity Through
    Cosmology}, Nuovo Cim. \textbf{122B}, 135-155 (2007),
    \texttt{arXiv:gr-qc/0702030}.
  \bibitem{lqg} 
    C.~Rovelli, {\em Quantum Gravity} (CUP, Cambridge, 2004);\\
    A.~Ashtekar and J.~Lewandowski, {Background Independent Quantum
    Gravity: A Status Report}, Class.Quant.Grav. \textbf{21}, R53 
    (2004), \texttt{arXiv:gr-qc/0404018};\\
    T.~Thiemann, {\em Introduction to Modern Canonical
    Quantum General Relativity} (CUP, Cambridge, 2007).
  \bibitem{aps} A.~Ashtekar, T.~Pawlowski and P.~Singh, {Quantum nature
    of the big bang}, Phys.Rev.Lett. \textbf{96}, 141301 (2006),
    \texttt{arXiv:gr-qc/0602086}.
  \bibitem{aps-old} A.~Ashtekar, T.~Pawlowski and P.~Singh, {Quantum nature
    of the big bang: An analytical and numerical investigation},
    Phys.Rev. {\bf D73}, 124038 (2006), \texttt{arXiv:gr-qc/0604013}.
  \bibitem{aps-imp} A.~Ashtekar, T.~Pawlowski and P.~Singh, {Quantum nature
    of the big bang: Improved dynamics}, Phys.Rev. {\bf D74}, 084003
    (2006), \texttt{arXiv:gr-qc/0607039}.
  \bibitem{pv} P.~Singh and K.~Vandersloot,  {Non-Singular Bouncing
    Universes in Loop Quantum Cosmology}, Phys.Rev. \textbf{D72}
    084004, (2005), \texttt{arXiv:gr-qc/0606032}. 
  \bibitem{apsv} A.~Ashtekar, T.~Pawlowski, P.~Singh and
    K.~Vandersloot, {Loop quantum cosmology of k=1 FRW models},
    Phys.Rev. {\bf D75}, 024035 (2007), \texttt{arXiv:gr-qc/0612104}.
  \bibitem{klp} L.~Szulc, W.~Kaminski and J.~Lewandowski, 
    {Closed FRW model in Loop Quantum Cosmology}, Class.Quant.Grav. 
    \textbf{24}, 2621-2635 (2007), \texttt{arXiv:gr-qc/0612101}.
  \bibitem{klp-nL} W.~Kaminski and J.~Lewandowski, 
    {The flat FRW model in LQC: the self-adjointness}, (2007),
    \texttt{arXiv:0709.3120}.
  \bibitem{acs} A.~Ashtekar, A.~Corichi and P.~Singh, {Robustness of
    key features of loop quantum cosmology}, (2007),
    \texttt{arXiv:0710.3565}. 
  \bibitem{cs-rec} A.~Corichi, P.~Singh, {Quantum bounce and cosmic
    recall}, (2007), \texttt{arXiv:0710.4543}.
  \bibitem{chv} D.~W.~Chiou, {Loop Quantum Cosmology in Bianchi Type I
    Models: Analytical Investigation}, Phys.Rev. \textbf{D75}, 024029
    (2007), \texttt{arXiv:gr-qc/0609029};\\
    D.~W.~Chiou, {Effective Dynamics for the Cosmological Bounces in
      Bianchi Type I Loop Quantum Cosmology}, (2007),
    \texttt{arXiv:gr-qc/0703010};\\
    D.~W.~Chiou and K.~Vandersloot, {The behavior of non-linear
      anisotropies in bouncing Bianchi I models of loop quantum
      cosmology}, Phys.Rev. \textbf{D76}, 084015 (2007),
    \texttt{arXiv:0707.2548};\\
    D.~W.~Chiou, {Effective Dynamics, Big Bounces and Scaling Symmetry
      in Bianchi Type I Loop Quantum Cosmology},
    Phys.Rev. \textbf{D76}, 
    124037 (2007), \texttt{arXiv:0710.0416};\\
    L.~Szulc, {Loop Quantum Cosmology of Diagonal Bianchi Type I
      model: simplified theory}, (2008), \texttt{arXiv:0803.3559}.
  \bibitem{gp} M.~Campiglia, R.~Gambini and J.~Pullin, {Loop
    quantization of spherically symmetric midi-superspaces},
    Class.Quant.Grav. \textbf{24}, 3649-3672 (2007),
    \texttt{arXiv:gr-qc/0703135}. 
  \bibitem{gm-letter} M.~Mart\'in-Benito, L.~J.~Garay and G.~A.~Mena 
    Marug\'an, {Hybrid Quantum Gowdy Cosmology: Combining Loop and Fock
    Quantization}, (2008), \texttt{arXiv:0804.1098}.
  \bibitem{eff} J.~Mielczarek, T.~Stachowiak, M.~Szyd{\l}owski, {Exact
    solutions for Big Bounce in loop quantum cosmology}, (2008),
    \texttt{arXiv:0801.0502}.
  \bibitem{abl} A.~Ashtekar, M.~Bojowald, J.~Lewandowski,
    {Mathematical structure of loop quantum cosmology}, Adv.Theo.%
    Math.Phys. \textbf{7}, 233-268 (2003), \texttt{arXiv:gr-qc/0304074}. 
  \bibitem{lqc-MB}
    M.~Bojowald, {Absence of singularity in loop quantum cosmology},
    Phys.Rev.Lett. \textbf{86}, 5227-5230 (2001),
    \texttt{arXiv:gr-qc/0102069};\\ 
    M.~Bojowald, {Isotropic loop quantum cosmology},
    Class.Quant.Grav. \textbf{19}, 2717-2741 (2002), 
    \texttt{arXiv:gr-qc/0202077}.
  \bibitem{ThTrick} T.~Thiemann, Anomaly-free formulation of
    non-perturbative, four-dimensional Lorentzian quantum gravity,
    {Phys.Lett.} \textbf{B380}, 257-264 (1998),
    \texttt{gr-qc/9606088};\\
    T.~Thiemann, Quantum spin dynamics (QSD), 
    {Class.Quant.Grav.} \textbf{15}, 839-873 (1998),
    \texttt{gr-qc/9606089};\\ 
    T.~Thiemann, QSD V : Quantum gravity as the natural regulator of
    matter quantum field theories, {Class.Quant.Grav.} \textbf{15}, 
    1281-1314 (1998), \texttt{gr-qc/9705019}.
  \bibitem{gave} 
    D.~Marolf, Refined algebraic quantization: Systems with a single
    constraint, (1995), \texttt{arXiv:gr-qc/9508015};\\
    D.~Marolf, {Quantum observables and recollapsing dynamics}, 
    Class.Quant.Grav. {\bf 12}, 1199-1220 (1995), 
    \texttt{arXiv:gr-qc/9404053};\\
    D.~Marolf, {Observables and a Hilbert space for Bianchi IX}, 
    Class.Quant.Grav. {\bf 12}, 1441-1454 (1995),
    \texttt{arXiv:gr-qc/9409049};\\
    D.~Marolf, {Almost ideal clocks in quantum cosmology: A brief
      derivation of time}, Class.Quant.Grav. {\bf 12}, 2469-2486 (1995),
    \texttt{arXiv:gr-qc/9412016};\\
    A.~Ashtekar, J.~Lewandowski, D.~Marolf,
    J.~Mour\~ao and T.~Thiemann, Quantization of diffeomorphism
    invariant theories of connections with local degrees of freedom,
    {J.Math.Phys.} \textbf{36}, 6456-6493 (1995), 
    \texttt{arXiv:gr-qc/9504018}.
  \bibitem{BesK} A.~Gil, J.~Segura and N.~M.~Temme, Evaluation of the
    modified Bessel function of the third kind of imaginary orders, {
      J.Comput.Phys.} {\bf 175}, 398-411 (2002);\\
    A.~Gil, J.~Segura and N.~M.~Temme, Computation of the modified Bessel
    function of the third kind of imaginary orders: uniform Airy-type
    asymptotic expansion, {J.Comput.App.Math.} {\bf 153}, 225-234
    (2003);\\
    A.~Gil, J.~Segura and N.~M.~Temme, 
    Computing solutions of the modified Bessel differential equation for
    imaginary orders and positive arguments,
    {ACM Trans.Math.Soft.} {\bf 30}, 145-158 (2004),
    \texttt{arXiv:math/0401128};\\
    A.~Gil, J.~Segura and N.~M.~Temme, 
    Algorithm 831: Modified Bessel functions of imaginary order and
    positive argument, {ACM Trans.Math.Soft.} {\bf 30}, 159-164
    (2004), \texttt{arXiv:cs/0401008}.
  \bibitem{mb-L} M.~Bojowald, {Harmonic cosmology: How much can we
    know about a universe before the big bang?}, (2007),
    \texttt{arXiv:0710.4919v1};\\
    M.~Bojowald, {Quantum nature of cosmological
    bounces}, (2008), \texttt{arXiv:0801.4001}.
  \bibitem{kp-rec} W.~Kami\'nski and T.~Pawlowski, (2008), 
    {\it in preparation}
  \bibitem{VT} V.~Taveras, IGPG preprint, (2006).
  \bibitem{GQM} A.~Ashtekar and T.~Schilling, Geometrical formulation of
    quantum mechanics, In: \textit{On Einstein's path}, A.~Harvery, ed
    (Springer-Verlag, New York, 1998), \texttt{arXiv:gr-qc/9706069}.
    T.~Schilling, Geometry of quantum mechanics, Ph.D. Dissertation,
    Penn State (1996), \texttt{%
    http://cgpg.gravity.psu.edu/archives/thesis/index.shtml}
\end{thebibliography}
\end{document}